\documentclass[letterpaper,12pt]{article}
\usepackage[T1]{fontenc}
\usepackage{bm}
\usepackage{amsmath}
\usepackage{amssymb}
\usepackage{amsfonts, mathrsfs}
\usepackage{eucal}
\usepackage{bmpsize}
\usepackage{graphicx}
\usepackage[authoryear]{natbib}
\usepackage[unicode=true,
 bookmarks=false,
 breaklinks=false,pdfborder={0 0 1},backref=false,colorlinks=false]
 {hyperref}

\makeatletter

\providecommand{\tabularnewline}{\\}

\@ifundefined{date}{}{\date{}}
\allowdisplaybreaks

\usepackage{latexsym}
\usepackage{multirow}
\usepackage{url}% not crucial - just used below for the URL 
\usepackage{enumerate}
\usepackage{color, colortbl}
\usepackage[table]{xcolor}
\usepackage{mathtools}
\usepackage{array}
\usepackage{footnote}
\usepackage{tikz}
\usetikzlibrary{fit,positioning}
\usepackage{booktabs}
\usepackage{upgreek}

\definecolor{Gray}{gray}{0.9}
\def\bSig\mathbf{\Sigma}

\newcommand{\bg}{\bm{g}}
\newcommand{\blam}{\bm{\lambda}}
\newcommand{\tildelta}{\widetilde{\delta}}

\usepackage[colorinlistoftodos,textwidth=2.1cm]{todonotes}

 \newcommand{\indep}{\perp\!\!\!\!\perp}

\usepackage{colortbl}
\usepackage{subcaption}

\usepackage{dcolumn}
\newcolumntype{.}{D{.}{.}{-1}}
\newcolumntype{d}[1]{D{.}{.}{#1}}
\usepackage{theorem}
\theoremstyle{definition}
\newtheorem{assumption}{Assumption}

\usepackage{rotating}

\usepackage{arydshln}

\usepackage[compact]{titlesec}

\allowdisplaybreaks

\newcommand{\spacingset}[1]{\renewcommand{\baselinestretch}%
{#1}\small\normalsize}

\makeatother

\begin{document}
\title{\Large\textbf{Transporting survival of an HIV clinical trial to the external target populations}}
\author
{\small\vspace{1ex} \textbf{Dasom Lee, Sujit Ghosh, Shu Yang$^{*}$}\\
\small \vspace{1ex}Department of Statistics, North Carolina State University, Raleigh, NC, U.S.A. \\
\small\vspace{1ex}$^{*}$\textit{email:} syang24@ncsu.edu}

\maketitle

\spacingset{1.3} 
\begin{abstract}
Due to the heterogeneity of the randomized controlled trial (RCT) and external target populations, the estimated treatment effect from the RCT is not directly applicable to the target population. 
For example, the patient characteristics of the ACTG 175 HIV trial are significantly different from that of the three external target populations of interest: US early-stage HIV patients, Thailand HIV patients, and southern Ethiopia HIV patients.
This paper considers several methods to transport the treatment effect from the ACTG 175 HIV trial to the target populations beyond the trial population. 
Most transport methods focus on continuous and binary outcomes; 
on the contrary, we derive and discuss several transport methods for survival outcomes: an outcome regression method based on a Cox proportional hazard (PH) model, an inverse probability weighting method based on the models for treatment assignment, sampling score, and censoring, and a doubly robust method that combines both methods, called the augmented calibration weighting (ACW) method.
However, as the PH assumption was found to be incorrect for the ACTG 175 trial, the methods that depend on the PH assumption may lead to the biased quantification of the treatment effect.
To account for the violation of the PH assumption, we extend the ACW method with the linear spline-based hazard regression model that does not require the PH assumption. 
Applying the aforementioned methods for transportability, we explore the effect of PH assumption, or the violation thereof, on transporting the survival results from the ACTG 175 trial to various external populations.

\end{abstract}
\newpage{}

\spacingset{1.5} 

\section{Introduction}\label{sec:intro trans}
In biomedical research, the distribution of baseline covariates in the randomized controlled trials (RCTs) sample, which is drawn from the \textit{trial population}, is often different from that of the \textit{target population}, the population that we want to make inferences about. Consequently, the estimated average treatment effect in the trial is not directly applicable to the target population beyond the trial population. 
As the findings from the RCTs often suffer from a lack of external validity \citep{rothwell2005external}, observational studies that include large samples that are representative of the target population have been widely used as a complement to the RCTs.
Several recent works have proposed methods of extending RCT findings to the target population by balancing the distribution of the baseline covariates between the RCT and the observational data.
A popular approach for covariate balancing is through direct modeling of the probability of participating in the trial, known as inverse probability of sampling weighting (IPSW) \citep{cole2010generalizing, stuart2011use, westreich2017transportability}. However, the IPSW approach is unstable under extreme sampling scores. 
Alternatively, the entropy balancing or calibration weighting approach has been used to enforce the covariate balance between the RCT and observational study without explicit modeling the sampling scores \citep{josey2021transporting,lee2021improving,lee2022generalizable}.
%using both the RCTs and the observational studies. 
There have been several terms for describing the process of extending the trial findings.
The problem of transporting the results to the external target population beyond the trial has been termed  \textit{transportability} \citep{pearl2011transportability,rudolph2017robust,westreich2017transportability,dahabreh2019extending,josey2021transporting}. Here the external target population is defined as the population that consists of at least some trial-ineligible individuals \citep{dahabreh2020extending}.
A similar problem that aims to generalize the findings from the RCT to its larger population has been termed  \textit{generalizability} \citep{cole2010generalizing, tipton2013improving, dahabreh2019generalizing, lee2021improving}. 
Both the generalizability and transportability try to balance the distribution of the baseline covariates between the RCT and the observational data.
The subtle differences between transportability and generalizability have been discussed in Lee et al. \citep{lee2021improving} and Colnet et al \citep{colnet2020causal}. 
%In this paper, we focus on the transportability problem. \dl{(The description of the transportability problem was already discussed in four lines above... need more?)}

% A popular approach for balancing the RCT and the observational sample is through direct modeling of the probability of participating in the trial, known as inverse probability of sampling weighting (IPSW) \citep{cole2010generalizing, stuart2011use, westreich2017transportability}. However, the IPSW approach is known to be unstable under extreme sampling scores. Alternatively, entropy balancing or calibration weighting approach has been used to enforce the covariate balance between the RCT and observational study without explicit modeling of the sampling scores \citep{josey2021transporting,lee2021improving,lee2022generalizable}. Colnet et al. \citep{colnet2020causal} presented a comprehensive review of various methods for generalizing and transporting trial results to the target population.

In this paper, we are interested in transporting the treatment effect for survival in the AIDS Clinical Trials Group (ACTG) 175  trial of intermediate-stage disease patients in the US.
We consider three external target populations - US early-stage HIV patients, Thailand HIV patients, and southern Ethiopia HIV patients - whose patient characteristics differ significantly from that of the ACTG 175 trial.
As a result, the treatment effect in the ACTG 175 HIV trial and that in the target populations are likely to be different.
To evaluate the treatment effect in the target populations, we consider several methods for transportability. 
Specifically, we consider two intuitive approaches, the outcome regression approach for survival outcomes that is similar to the one proposed by Chen and Tsiasis \citep{chen2001causal} under the PH assumption and the inverse probability weighting (IPW) approach that involves treatment propensity score, calibration weighting, and censoring model. 
We also consider a semiparametric efficient estimator, called the augmented calibration weighting (ACW) estimator, proposed by Lee et al. \citep{lee2022generalizable}. 
The ACW estimator combines the two intuitive approaches under a semiparamteric theory \citep{tsiatis2006semiparametric}. 
It accounts for the heterogeneity between the RCT and the observational study for a broad class of survival estimands that are functionals of treatment-specific survival functions, such as a difference in survival probabilities and restricted mean survival times. 
The ACW estimator is doubly robust in the sense that it is a consistent estimator if the Cox PH model for the survival outcome is correctly specified or the weighting models are correctly specified, and is the most efficient estimator when all models are correctly specified.

%As the comparison of the distribution of treatment-specific survival times is often of interest in clinical or biomedical studies, Lee et al. \citep{lee2022generalizable} proposed a semiparametric efficient estimator, called augmented calibration weighting (ACW) estimator, which accounts for the heterogeneity between the RCT and the observational study for a broad class of survival estimands that are functionals of treatment-specific survival functions, such as difference in survival probability and restricted mean survival times. 
%Their approach combines the Cox proportional hazard (PH) regression model for the survival time and inverse probability weighting (IPW) approach that involves treatment propensity score, calibration weighting, and censoring model. 
%The ACW estimator is doubly robust in the sense that it is consistent if the Cox PH model or the weighting models are correctly specified. 
However, the proportionality assumption on which the ACW estimator depends may be questionable in many clinical trials in practice, particularly when there is a delayed effect or a crossing of the survival functions. 
Sheng and Ghosh \citep{sheng2020effects} explored various PH and non-PH models, including the linear spline-based hazard regression (HARE) model. 
They found that the PH assumption may be violated in the ACTG 175 trial data and showed that the violation of the PH assumption results in the biased variable section for the PH models but not for the non-PH models.
%, via simulation studies and an application to the ACTG 175 trial data where the PH assumption may be violated.
Consequently, the methods that depend on the PH assumption may lead to the biased quantification of the treatment effect in the target populations.
To account for this possible violation of the PH assumption, we extend the ACW estimator with the HARE model that does not require the PH assumption. 
This extension enhances the robustness of the ACW estimator compared to the original counterpart.
%The resulting estimator is robust to the violation of the PH assumption whereas the original ACW estimator may not.
We apply the aforementioned methods for transportability to explore the effect of the PH assumption, or the violation thereof, on transporting the treatment effect from the ACTG 175 trial to the target populations.

%Applying the aforementioned methods for transportability, we explore the effect of PH assumption, or the violation thereof, on transporting the results from the ACTG 175 trial.

%In this paper, we study the effect of the PH assumption when transporting the treatment effect in the RCT to the target population through a comparative case study. 
%Specifically, we consider two intuitive approaches, outcome regression approach for survival outcomes that similar to the one proposed by Chen and Tsiasis \citep{chen2001causal} under the PH assumption and the IPW approach, and the ACW estimator which is a combination of them. 
%The ACW estimator is doubly robust in the sense that it is a consistent estimator if the Cox PH model for the survival outcome is correctly specified or the weighting models are correctly specified, and is the most efficient estimator when all models are correctly specified under semiparametric theory.
%Moreover, to account for the possible violation of the PH assumption, we extend the ACW estimator with HARE model that does not require the PH assumption. 
%Consequently, the resulting estimator is robust to the violation of the PH assumption whereas the original ACW estimator may not. We apply the aforementioned methods to transport the treatment effect estimated from the ACTG 175 trial to three external target populations - US early-stage HIV patients, Thailand HIV patients, and southern Ethiopia HIV patients - beyond the trial population which is the intermediate-stage HIV patients in the US.

The remainder of the paper is organized as follows. Section \ref{sec:motive} provides the details of the motivating example, the ACTG-175 trial and three external datasets.
Section \ref{sec:notandassum} formalizes the basic causal inference framework and discusses identifiability conditions for transportability.
Section \ref{sec:trans method} presents an overview of PH and non-PH models for transportability. 
In Section \ref{sec:trans data}, we present the result of transporting the treatment effect to three target populations. In Section \ref{sec:conclude}, we provide discussion and concluding remarks.

\section{A motivating example: ACTG-175 trial and external data}\label{sec:motive}

%\subsection{ACTG 175 trial and observational studies} \label{sec:actg}
To explore the effect of PH assumption on transportability problem, we consider an HIV clinical trial.
The ACTG 175 trial enrolled HIV-infected patients with 200 - 500 cells/mm$^3$ CD4 cell counts who were randomized to four antiretroviral therapies: Zidovudine monotherapy (ZDV), Zidovudine plus Didanosine (ZDV + ddI), Zidovudine plus Zalcitabine (ZDV + ZAL), or Didanosine monotherapy (ddI) \citep{hammer1996trial}. For illustration purposes, we choose ZDV + ddI and ZDV as binary treatment, which consists of 522 ZDV + ddI patients and 532 ZDV monotherapy patients. 
The analyses for the effect of two other treatments, ZDV + ZAL and ddI, over ZDV, are provided in Appendix \ref{app:other trts}.
The primary endpoint of the study is the progression of HIV disease, defined as a more than 50 percent decline in the CD4 cell count or development of the acquired immunodeficiency syndrome, or death. The causal estimand of interest is a 2-year event-free survival difference between ZDV + ddI and ZDV monotherapy, as at least 24 months of follow-up is required for the ACTG 175 trial. About 73\% of the survival times were right-censored.

The ACTG 175 trial data are available from the R package \texttt{speff2trial} and were previously used in Yang et al. \citep{yang2021smim} and Sheng and Ghosh \citep{sheng2020effects}; all their analyses focus on comparing the effect of combination therapy and the ZDV monotherapy in the trial population.
Responses to treatment, however, may vary according to the patient characteristics such as disease stage and the history of prior drug exposure \citep{hammer1996trial}.
Kennedy et al. \citep{kennedy2021semiparametric} have found that combination therapy may be more effective for the high-risk patients, but its utility for low-risk patients remains unclear.
Therefore, our interest lies in transporting the results from the ACTG 175 trial with intermediate-stage HIV patients to the external target populations with the higher- and lower-risk compared to trial patients to evaluate the treatment effect in the target population. 
Specifically, we consider the following three target populations: 
\begin{enumerate}
    \item  US early-stage HIV patients represented by the observational Acute Infection and
Early Disease Research Program (AIEDRP) database \citep{hecht2006multicenter}
    \item HIV patients in Thailand represented by the retrospective study by Manosuthi et al. \citep{manosuthi2021retrospective}
    \item HIV patients in southern Ethiopia represented by the retrospective study by Hailemariam et al. \citep{hailemariam2016determinants}
\end{enumerate}

The baseline characteristics of the ACTG 175 trial and the three external datasets are summarized in Table \ref{tab:baseline}. We select seven covariates as these were considered prognostic factors for disease progression following the previous research \citep{sheng2020effects} and also based on the external datasets availability. 
Even though not all factors are available in the external observational data as these were not collected for research purposes, some important covariates that are predictive biomarkers, e.g., CD4 count and drug, have been captured in external datasets.
%Note that since observational samples were not collected for research purposes, not all important factors may be available in the observational data. 
As seen in Table \ref{tab:baseline}, patients in the ACTG 175 trial are significantly different from those in the external datasets. Specifically, AIEDRP data consists of patients with higher CD4 counts and less history of previous intravenous drugs. On the other hand, patients in Thailand and southern Ethiopia studies have significantly lower CD4 count compared to the ACTG 175 trial. Thus, transporting to the external populations will lead to different quantification of the treatment effect.

\begin{figure}[ht]
\centerline{
\includegraphics[width=7in]{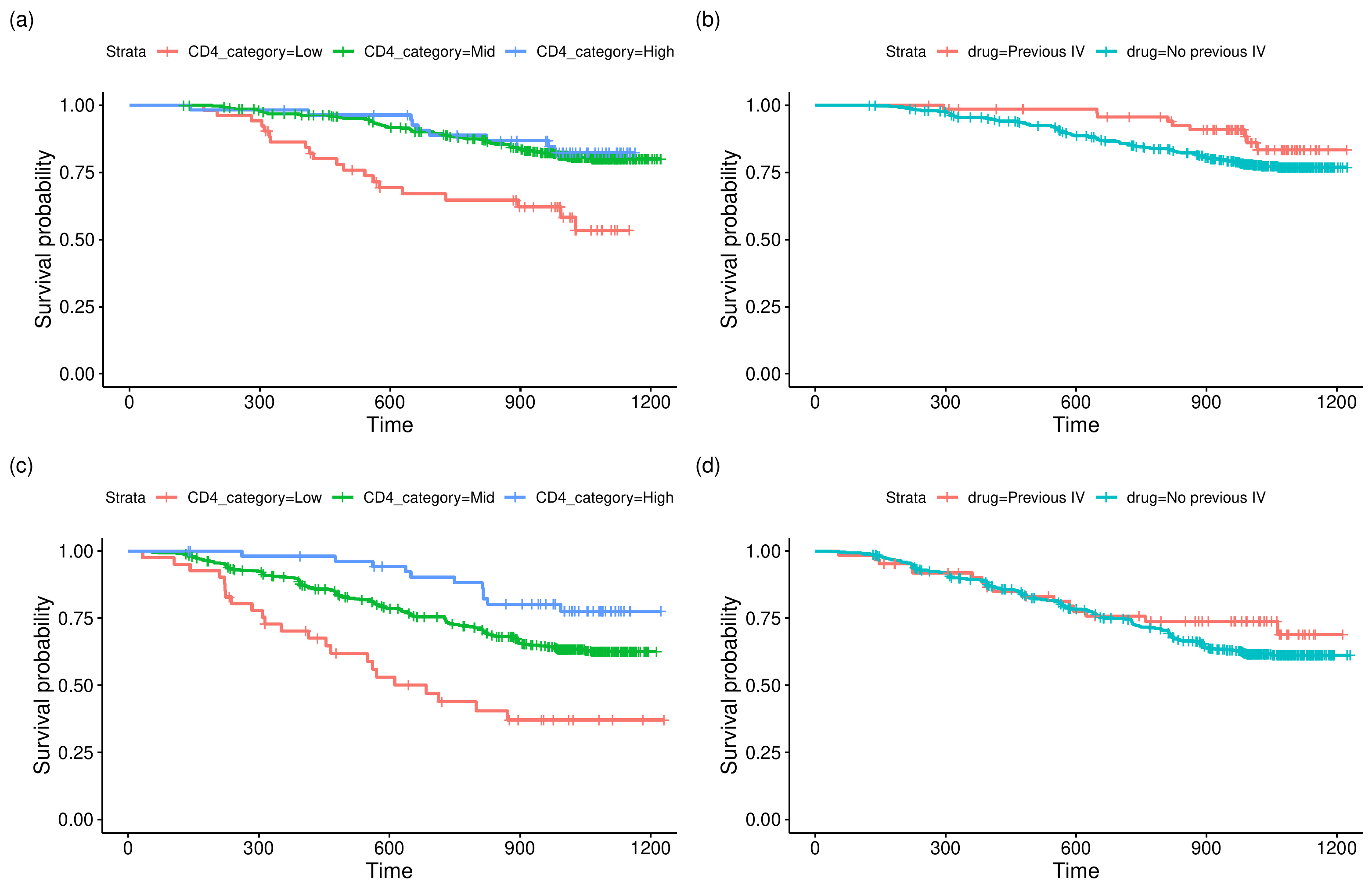}}
\caption{Treatment-specific Kaplan-Meier (KM) curves by two prognostic covariates; (a) KM curves of ZDV + ddI treatment group by CD4 category; (b) KM curves of ZDV + ddI treatment group by drug; (c) KM curves of ZDV treatment group by CD4 category; (d) KM curves of ZDV treatment group by drug. The CD4 category is defined as in Table \ref{tab:baseline}. \label{fig:surv ph viol}}
\end{figure}

\begin{center}
\begin{table}[ht]%
\centering
\caption{Summary of baseline characteristics of the ACTG 175 trial and the three external datasets.\label{tab:baseline}}
        \resizebox{\textwidth}{!}{%
\begin{tabular}{lcccc}
\toprule
& & \multicolumn{3}{c}{\textbf{External Datasets}} \tabularnewline\cmidrule{3-5}
 & \textbf{ACTG 175 Trial}  & \textbf{US Early-Stage}  & \textbf{Thailand}  & \textbf{Southern Ethiopia}\\
\textbf{Variable} & (n = 1054) & (n = 1762) & (n = 11911) & (n = 2579) \tabularnewline
\midrule
 \textbf{Male}, \% & 82.16  & 95.46  & 67.7  & 44.5   \tabularnewline
 \rowcolor{Gray} \textbf{Age} (year), Mean $\pm$ SD & 35.23 $\pm$ 8.77  & 34.99 $\pm$ 8.48  & 32 $\pm$ 11 & 32.5 $\pm$ 9.1  \tabularnewline
 \textbf{CD4 count} (cells/mm$^3$), Mean $\pm$ SD & 351 $\pm$ 122.3 & 545.7 $\pm$ 228.3 &  & 164.02 $\pm$ 117.84 \tabularnewline
 \rowcolor{Gray} \textbf{CD4 category}, \% & & & & \tabularnewline
 \rowcolor{Gray} \ \ \ Low ($\le$ 200) & 8.82 & 2.33 & 52.1 & \tabularnewline
 \rowcolor{Gray} \ \ \ Mid (201 - 500)& 80.65 & 45.63 & 13.7 & \tabularnewline
 \rowcolor{Gray} \ \ \ High ($>$ 500) & 10.53 & 52.04 & 3.6 & \tabularnewline
 \textbf{Race} (White), \% & 72.11 & 67.14 & \tabularnewline
 \rowcolor{Gray} \textbf{Drug}, \% & 12.9 & 3.92 & & \tabularnewline
 \textbf{Weight} (kg), Mean $\pm$ SD & 75.47 $\pm$ 13.43 &  & & 52.22 $\pm$ 9.13  \tabularnewline
\bottomrule
\end{tabular}
}
\end{table}
\end{center}
Although the trial and external datasets provide a unique opportunity to understand how treatment works in large target populations, they also present two major challenges that need to be addressed, one intrinsic to the trial data and the other one with respect to the external data. First, for the ACTG 175 trial, there is a possibility that the common PH assumption is violated. 
Figure \ref{fig:surv ph viol} depicts the treatment-specific Kaplan-Meier (KM) curves by two prognostic covariates, i.e., CD4 category and drug. 
Instead of CD4 count, we use CD4 category defined in Table \ref{tab:baseline} to illustrate via KM curves. Figure \ref{fig:surv ph viol}(a) and Figure \ref{fig:surv ph viol}(d) depict the possible violation of the PH assumption by the delayed effect and the crossing curves. 
Figure \ref{fig:surv ph viol}(b) and Figure \ref{fig:surv ph viol}(c) also depict the possible violation by the delayed and decreasing effects. 
To formally test the validity of the PH assumption, similar to Sheng and Ghosh \citep{sheng2020effects}, we checked the relationship between the Schoenfeld residuals and the time\citep{grambsch1994proportional} using the R function \textit{cox.zph}. We use the KM transformed time, i.e., $\widehat{S}(t) = \prod_{i:t_i \le t}(1 - d_i/n_i)$, where $t_i$ is a time with at least one event happened, $d_i$ is the number of events happened at $t_i$, and $n_i$ is the number of subjects survived up to $t_i$, which is a default option to avoid the extreme outlier and for efficiency. For the ZDV + ddI treatment group, the global $p$-value is 0.028, implying that the null hypothesis of no departure from proportionality is rejected. In particular, covariates that show significant violation of proportionality are CD4 count ($p$ = 0.0092) and drug ($p$ = 0.033). For the ZDV monotherapy group, the global test is not significant (p = 0.194), but there is a possibility of departure from proportionality for CD4 count (p.= 0.049).
Figure \ref{fig:surv ph viol} and the test using the Schoenfeld residuals suggest the possible PH violation in the ACTG 175 trial. Thus, using a Cox PH model without adjusting for such violation could lead to a biased estimation of the treatment effect in the trial as well as in the target populations. This also suggests the necessity of non-PH models that do not require the PH assumption. 
%In the following sections, we 

%In the following sections, we compare the treatment effects between ZDV + ddI and ZDV monotherapy in HIV patients in aforementioned three external populations transporting the results from the ACTG 175 trial. In addition to the methods for transportability described in Section \ref{sec:trans method}, we compare the results from the Cox PH model and the HARE model only based on the ACTG 175 trial.

%the existing approach did not address. 
Second, for the external observational data for Thailand and southern Ethiopia, we were only able to find the summary statistics. As individual-level data were not publicly available, we emulated the external datasets based on the summary statistics incorporating the correlation structure of the trial data. 
The details of how we emulated the data are described in Section \ref{sec:thai_and_etio}. 
Moreover, some important covariates were captured only in a few external datasets or in different data types. 
For example, the covariate drug was collected only in the US early-stage data, and the summary of the baseline covariate CD4 count is provided in categories for the Thailand study.

\section{Notation and Identification for Transportability}\label{sec:notandassum}

We now introduce some basic notations and identifiability conditions to transport the findings from the RCT, i.e., the ACTG 175 trial, to the target populations that are represented by three external datasets.
Let $X$ be a vector of $p$-dimensional baseline covariates and $A$ be the binary treatment, $A = \{0, 1\}$. Employing the potential outcomes framework \citep{rubin1974estimating, rubin1986comment}, let $T^a$ be the potential survival time if a subject receives the treatment $A = a$. The survival time $T$ is defined as $T = T^1A + T^0(1-A)$. Let $C$ be the censoring time. We assume noninformative censoring conditional on covariates and treatment throughout the paper. Due to the censoring, the survival time $T$ is not always observable for all subjects. 
Instead, we observe $U = \min(T, C)$ and $\Delta = I(T \ge C)$ where $I(\cdot)$ is an indicator function.

In transportability, we consider a super-population framework in which the trial and target populations are subpopulations. 
The RCT is sampled from a trial population with an unknown mechanism, and the external observational sample is randomly sampled from the external target population with some known mechanism. 
Let $\delta = 1$ denote the binary indicator of trial participation and $\tildelta = 1$ denote the binary indicator of external observational study participation. From the RCT sample, we observe $\{U_i, \Delta_i, A_i, X_i, \delta_i = 1, \widetilde{\delta}_i = 0\}_{i=1}^n$ from $n$ independent and identically distributed subjects. For the external data, we consider a common setting that only the covariates information is available, i.e., $\{X_i, \delta_i = 0, \widetilde{\delta}_i = 1\}_{i=n+1}^{n+m}$, from $m$ independent and identically distributed subjects drawn from the target population \citep{dahabreh2019extending, lee2022generalizable}.

When comparing the survival of the two treatments, a popular estimand of interest is the difference in survival probability at landmark times. Let $S_a(t) = P(T^a \ge t)$ be the treatment-specific survival probability curve, $a \in \{0, 1\}$. Accordingly, we define the target average treatment effect (TATE) \citep{josey2021transporting}: $\tau = E\{I(T^1 \ge t) - I(T^0 \ge t) \mid \tildelta = 1\} = S_1(t\mid \tildelta = 1) - S_0(t\mid \tildelta = 1)$.

Define the treatment propensity score $\pi_{A}(X) = P(A = 1 \mid X, \delta = 1)$ and the sampling score $\pi_{\delta}(X) = P(\delta= 1 \mid X)$. We now discuss the conditions to identify the TATE.

\begin{assumption}[Ignorability and positivity of trial treatment assignment]\label{assump:trt ignorability}
\textcolor{white}{1}\\
(i) $\{T^0, T^1\} \indep A\mid(X,\delta=1)$; and (ii) $0<\pi_{A}(X)<1$ with probability $1$.
\end{assumption}

\begin{assumption}[Conditional survival transportability]\label{assump:sirv trans}
\textcolor{white}{1}\\
$P(T^a \ge t \mid X, \delta = 1) = P(T^a \ge t \mid X, \tildelta = 1)$ for $\forall x$ s.t. $P(X = x \mid \tildelta = 1) > 0$ and $a \in \{0, 1 \}$.
\end{assumption}

\begin{assumption}[Positivity of trial participation]\label{assump:trial positivity} 
$0 < \pi_{\delta}(X) < 1$ for $\forall x$ s.t. $P(X = x \mid \tildelta = 1) > 0$.
\end{assumption}
\noindent
Assumption \ref{assump:trt ignorability} is a standard assumption and is likely to hold for well-defined RCTs in general.
As the ACTG 175 trial implemented treatment randomization and had good patient compliance, Assumption \ref{assump:trt ignorability} is valid.
Assumptions \ref{assump:sirv trans} and \ref{assump:trial positivity} are needed to transport findings from the RCT to the target population. 
Assumption \ref{assump:sirv trans} is formally weaker than the ignorability assumption on trial participation \citep{stuart2011use, westreich2017transportability}, i.e., $\{T^0, T^1\} \indep \delta \mid X$, but it suffices to identify the TATE as well as the potential survival curves in the target population, $P(T^a \ge t \mid \tildelta = 1)$. 
Several previous literature discussed the weaker version of Assumption \ref{assump:sirv trans}, $E\{I(T^1 \ge t) - I(T^0 \ge t) \mid X, \delta = 1\} = E\{I(T^1 \ge t) - I(T^0 \ge t) \mid X, \tildelta = 1\}$, called exchangeability in measure\citep{dahabreh2020toward,josey2021transporting}. 
However, the potential survival curves in the target population are not identifiable under the weaker version. 
Assumption \ref{assump:sirv trans} is not verifiable but plausible as some important covariates that are predictive of the disease progression, e.g., CD4 count and drug, were captured both in the ACTG 175 trial and external datasets.
Assumption \ref{assump:trial positivity} implies the absence of patient characteristics in the target population that prevent them from participating in the trial; thus, covariates separating the trial and the target population should be excluded \citep{tipton2013improving,dahabreh2020extending}, which is valid for the ACTG 175 trial and external data we are interested in.
%Assumptions \ref{assump:trt ignorability} -- \ref{assump:trial positivity} are not testable in general, and their plausibility should be justified by subject matter knowledge and sensitivity analyses. 

In addition, we assume that the external observational data is drawn from the target population, can be under a simple random sampling or more complex sampling designs. Accordingly, we define a known sampling weight $d$.

\begin{assumption}[The known design weight for the external observational sample]\label{assump:knownOX}\textcolor{white}{1}\\
The observational sample design weight $d = 1/P(\widetilde{\delta} = 1 \mid X)$ is known.
\end{assumption}
\noindent
Under the above assumptions, the TATE, or $S_a(t \mid \tildelta = 1)$, are identified using only the observed data, by 
\begin{align}
    S_a(t \mid \tildelta = 1) = \mathbb{E} \left\{S_{a}(t, X) \mid \tildelta = 1 \right\},
    \label{eq:ident-outcome}
\end{align}
where $S_{a}(t, X) = \mathbb{E}\{I(T \ge t) \mid X, A =a, \delta = 1,  C \ge t\}$ is the conditional survival curves, for $a \in \{0, 1\}$. Alternatively, $S_a(t \mid \tildelta = 1)$ can also be identified based on the IPW-based approach by
\begin{align}
\label{eq:ident-weighting}
S_a(t \mid \tildelta = 1) = \frac1{P(\tildelta = 1)}\mathbb{E}\left[\frac{\delta}{\pi_{\delta}(X)}\frac1{d}\frac{I(A=a)}{\pi_{A}(X)^a\{1-\pi_A(X)\}^{1-a} }\frac{Y(t)}{S_{a}^C(t, X)}\right],
\end{align}
where $Y(t) = I(U \ge t)$ and $S_a^C(t, X) = P(C > t \mid X, A = a, \delta = 1)$, the conditional censoring probability model.
By positivity assumptions, we have 
\[P(\tildelta = 1) = \mathbb{E}\left[\frac{\delta}{\pi_{\delta}(X)}\frac1{d}\frac{I(A=a)}{\pi_{A}(X)^a\{1-\pi_A(X)\}^{1-a} }\right].\]
Several methods for transportability motivated by each identification formula in \eqref{eq:ident-outcome} and \eqref{eq:ident-weighting} as well as jointly will be discussed in the following section.

\section{Estimation Methods for Transportability}\label{sec:trans method}

In this section, we present four different methods for transporting results from the ACTG 175 trial to estimate TATE under Assumptions \ref{assump:trt ignorability} --  \ref{assump:knownOX}. Due to the possible violation of the PH assumption in the ACTG 175 trial as described in Section \ref{sec:motive}, we consider both PH and non-PH models.
%Similar to Lee et al. \citep{lee2022generalizable}, these methods are motivated 

\subsection{Inverse Probability and Calibration Weighting (CW) Model}\label{sec:weighting}
A common approach for transportability is through the IPW-based approach following the identification formula in \eqref{eq:ident-weighting}. As shown in \eqref{eq:ident-weighting}, this approach can be viewed as a combination of IPSW,  inverse probability of treatment weighting (IPTW), and inverse probability of censoring weighting (IPCW).
With respect to sampling score, a key idea to account for the different patient characteristics between the RCT sample and the target population represented by the external data is to weight the RCT sample with the inverse odds of sampling \citep{westreich2017transportability}.
%, analogous to the IPSW that was introduced to generalize the trial result to its broad population \citep{cole2010generalizing, stuart2011use}. 
In particular, following IPSW, one can model $\pi_{\delta}(X) = \{{\omega}_{\text{IPSW}}(X)\}^{-1}$ in \eqref{eq:ident-weighting}, where 
 ${\omega}_{\text{IPSW}}(X) =  P(\tildelta = 1 \mid \delta + \tildelta = 1, X)/P(\delta = 1 \mid \delta + \tildelta = 1, X)$ using the common logistic regression model. However, the IPSW method may not be stable if $P(\delta = 1 \mid \delta + \tildelta = 1, X)$ is close to zero for some $X$, and it requires the sampling score model to be correctly specified.
 
 Instead of estimating the sampling scores directly, the calibration weighting (CW) approach has been used in many recent works \citep{josey2021transporting, lee2021improving, lee2022generalizable}, which is known to be more stable than the IPSW approach. The CW approach is analogous to the entropy balancing method \citep{hainmueller2012entropy} and is based on the following identity,
 \begin{equation}
E \left\{ \frac{\delta}{\pi_{\delta}(X)} \bg(X) \right\} = E \left \{\widetilde{\delta} d \bg(X) \right \} = E \{\bg(X) \mid \tildelta = 1\}, \label{eq:trans identity}
\end{equation}
where $\bg(X)$ is a function of $X$ to be calibrated, such as the moment functions or nonlinear transformations. That is, the calibrated RCT sample empirically matches the external data that represents the target population.
Empirically, the calibration weights $\omega_i$ are the solutions to the following optimization problem
\begin{align}
    \min_{\mathcal{W}}&\sum_{i=1}^n \omega_i\log \omega_i \label{eq:obj fn},\\
    \mbox{subject to } & \omega_i \ge 0, ~\forall i,  \sum_{i=1}^n \omega_i = 1, \mbox{ and } \sum_{i=1}^N\delta_i\omega_i \bg(X_i) = \widetilde{\bm{g}}, \notag
\end{align}
   where $\mathcal{W} = \{w_i: \delta_i = 1\}$ and $\widetilde{\bm{g}} = \sum_{i=1}^N\widetilde{\delta}_id_i\bg(X_i)/\sum_{i=1}^N\widetilde{\delta}_id_i$, a consistent estimator of $E\{\bg(X) \mid \tildelta = 1\}$.
  The calibration weights $\omega_i$ are also the solution to the Lagrangian dual problem $L(\blam,\mathcal{W})=\sum_{i=1}^{n}\omega_{i}\log \omega_{i}-\blam^{\top}\left\{ \sum_{i=1}^{n}\omega_{i}\bg(X_{i})-\widetilde{\bg}\right\}$ where $\widehat{\blam}$ solves $U(\blam)=\sum_{i=1}^{n}\exp\left\{ \blam^{\top}\bg(X_{i})\right\} \left\{ \bg(X_{i})-\widetilde{\bg}\right\} =0$. The estimated calibration weights are  $\widehat{\omega}_{i}=\omega(X_{i};\widehat{\blam})=\exp\{ \widehat{\blam}^{\top}\bg(X_{i})\} / [\sum_{i=1}^{n}\exp\{ \widehat{\blam}^{\top} \bg(X_{i})\} ]$. For the calibration weighting, Lee et al.\citep{lee2021improving} posited the loglinear sampling score model 
\begin{equation}
    \pi_{\delta}(X)=\exp\{\eta_{0}^{\mathrm{T}}\bg(X)\}, \mbox{ for some } \eta_{0}\label{eq:loglin sampling}.
\end{equation}
This is because the objective function in \eqref{eq:obj fn} has a solution that has the same functional form as inverse probability of sampling score under the loglinear model. It is known that $\widehat{\blam}$ is equivalent to $-\widehat{\eta}$ \citep{lee2021improving} thus $\widehat{\pi}_{\delta}(X)$ and $\widehat{\omega}_{i}$ can be expressed using $\widehat{\eta}$, i.e., $\widehat{\pi}_{\delta}(X)=\pi_{\delta}(X ; \widehat{\eta}) = \exp\{\widehat{\eta}^{\mathrm{T}}\bg(X)\}$ and  $\widehat{\omega}_{i}=\omega(X_{i};\widehat{\eta})=\exp\{ -\widehat{\eta}^{\top}\bg(X_{i})\} / [\sum_{i=1}^{n}\exp\{- \widehat{\eta}^{\top} \bg(X_{i})\}]$.

In addition to the sampling score model, one can model the treatment propensity scores using a common logistic regression model,
\begin{equation}
    \pi_{A}(X)=\left[1 + \exp\{-\rho_{0}^{\mathrm{T}}\bg(X)\} \right]^{-1}, \mbox{ for some } \rho_{0}\label{eq:propensity model}.
\end{equation}
Even though $\pi_A(X)$ is generally known for RCTs, many researchers have suggested estimating the treatment propensity scores which can increase the efficiency and account for the chance of imbalance of prognostic variables \citep[e.g.,][]{williamson2014variance, lee2022generalizable}. 
Moreover, the right censoring needs to be accounted for. We assume the noninformative censoring assumption conditional on covariates and treatment, i.e., $\{T^1, T^0\} \indep C \mid (X, A, \delta = 1)$, which also implies $T \indep C \mid (X, A, \delta = 1)$. A widely used Cox proportional hazard model can be used, with the conditional hazard 
\begin{align}
    \lambda^C(t \mid X, A = a) = \lambda^C_{a0}(t)\exp(\gamma_a^\mathrm{T}X), \mbox{for } a \in \{0, 1\}\label{eq:cox censoring}.
\end{align}

Define $\widehat{\pi}_{ai} = A_i\pi_{A}\left(X_i ; \widehat{\rho}\right) + (1-A_i)\left\{1-\pi_{A}\left(X_i ; \widehat{\rho}\right)\right\}$, $A_{ai} = I(A_i = a)$, and $\widehat{\Lambda}_{ai}^C(t) = \widehat{\Lambda}^C_{a0}(t) \exp(\widehat{\gamma}_a^{\mathrm{T}}X_i)$ where $\Lambda^C_{a0}(t) \equiv \int_0^t \lambda^C_{a0}(u)\rm{d}u$. The CW estimator of the treatment-specific survival curve is 

\begin{equation}
    \widehat{S}_a^{\text{CW}}(t) = \sum_{i=1}^N\delta_i\widehat{\omega}_i\frac{A_{ai}}{\widehat{\pi}_{ai}}e^{\widehat{\Lambda}_{ai}^C(t)}Y_{i}(t) 
\label{eq:cw trans}.
\end{equation} 
The IPSW estimator can be defined by replacing the calibration weights $\widehat{\omega}_i$ in \eqref{eq:cw trans} with $\widehat{\omega}_{\text{IPSW}}(X_i)$.

\subsection{Outcome Regression  with a PH assumption}\label{sec:coxph}
Another method for transporting beyond the trial population relies on the conditional survival curves in the identification formula \eqref{eq:ident-outcome}, known as the outcome regression (OR) approach. A widely used Cox PH model assumes the treatment-specific conditional hazard function as
\begin{equation}
    \lambda_{ai}(t) \equiv \lambda_a(t \mid X_i) = \lambda_{a0}(t)\exp(\beta_a^{\mathrm{T}}X_i) \label{eq:cox trans}.
\end{equation}
Following Chen and Tsiatis \citep{chen2001causal}, one can adjust for the imbalances between the RCT sample and the external data  by averaging the treatment effect in the trial sample over the external data. Specifically, first estimating the treatment-specific survival curve conditional on $\delta = 1$ under model \eqref{eq:cox trans}, i.e., $\widehat{S}_{a}(t, X_i) = \exp \left \{- \widehat{\Lambda}_{ai}(t)\right\} =  \exp \left \{- \widehat{\Lambda}_{a0}(t) \exp(\widehat{\beta}_a^{\mathrm{T}}X_i) \right \}$ for the RCT participants, then applying the design-weighted averaging over $f(X \mid \tildelta = 1)$ thus transporting the RCT results to the target population. 
The resulting OR estimator of the treatment-specific survival curve is 
\begin{align}
    \widehat{S}^{\text{OR}}_a(t) = & \left(\sum_{i=1}^{N} \tildelta_i \right)^{-1} \sum_{i=1}^N \tildelta_i  e^{-\widehat{\Lambda}_{ai}(t)},
    %= & m^{-1} \sum_{i=1}^N \tildelta_i e^{-\widehat{\Lambda}_{ai}(t)} \mbox{   in the simple random sampling},
    \label{eq:OR trans}
\end{align}
which is consistent under the PH assumption.
%\subsection{Inverse probability of sampling model}\label{sec:ipsw}

\subsection{Augmented Calibration Weighing (ACW) Model with a PH assumption}\label{sec:ACWsurv}
The CW estimator specified in \eqref{eq:cw trans} and the OR estimator specified in \eqref{eq:OR trans} are singly robust in that the former is consistent only under the correct weighting models \eqref{eq:loglin sampling} -- \eqref{eq:cox censoring} and the latter is consistent under the correct survival outcome regression model \eqref{eq:cox trans}.
Lee et al.\citep{lee2022generalizable} proposed an improved estimator, named the ACW estimator, that combines the CW and the OR estimators employing the semiparametric theory \citep{tsiatis2006semiparametric}. 
The ACW estimator is in the form of 
\begin{align}
  \widehat{S}^{\text{ACW}}_a(t) =  \exp \left \{-\int_0^t\frac{num}{denom} \right \},
  \label{eq:ACW trans}
 \end{align}
 where
 \begin{align}
  denom = & \sum_{i=1}^N \delta_i\widehat{\omega}_i\frac{A_{ai}}{\widehat{\pi}_{ai}} e^{\widehat{\Lambda}_{ai}^C(t)}Y_{i}(t) \notag\\
& + \sum_{i=1}^N e^{-\widehat{\Lambda}_{ai}(t)}  \left [\left(\sum_{i=1}^{N} \tildelta_i \right)^{-1}\tildelta_i - \delta_i\widehat{\omega}_i\frac{A_{ai}}{\widehat{\pi}_{ai}} \left \{1 - \int_0^t \left \{ e^{\widehat{\Lambda}_{ai}^C(u) + \widehat{\Lambda}_{ai}(u)} \right \}\mathrm{d}\widehat{M}_{ai}^C(u) \right \} \right ],
\label{eq:ACW denom}
\end{align}
and
 \begin{align*}
     num =& \sum_{i=1}^N \delta_i\widehat{\omega}_i\frac{A_{ai}}{\widehat{\pi}_{ai}} e^{\widehat{\Lambda}_{ai}^C(u)}\mathrm{d}N_{i}(u) \\
     &+ \sum_{i=1}^N e^{-\widehat{\Lambda}_{ai}(u)} \mathrm{d}\widehat{\Lambda}_{ai}(u) \left [ \left(\sum_{i=1}^{N} \tildelta_i \right)^{-1}\tildelta_i - \delta_i\widehat{\omega}_i\frac{A_{ai}}{\widehat{\pi}_{ai}} \left \{1 - \int_0^u \left \{ e^{\widehat{\Lambda}_{ai}^C(s) + \widehat{\Lambda}_{ai}(s)} \right \}\mathrm{d}\widehat{M}_{ai}^C(s) \right \} \right ].
 \end{align*}
This approach is based on the representation $\Lambda_a(t \mid \tildelta = 1) = \int_0^t-\{ {S}_a(u\mid \tildelta = 1)\}^{-1}{{\mathrm{d}S}_a(u\mid \tildelta = 1)}$ where $num$ 
 estimates $\mathrm{d}S_a(u\mid \tildelta = 1)$ and $denom$ estimates ${S}_a(u\mid \tildelta = 1)$ separately \citep{zhang2012contrasting}. 

The ACW estimator is a semiparametric efficient estimator that has the influence function with the smallest variance. The ACW estimator has two nice properties. First, it is doubly robust in the sense that it is a consistent estimator of $S_a(t)$ if the weighting models \eqref{eq:loglin sampling} -- \eqref{eq:cox censoring} are correctly specified or the outcome model \eqref{eq:cox trans} is correctly specified, not necessarily both. Also, the ACW estimator is locally efficient, i.e., it is the most efficient estimator when all working models are correctly specified.
Note that the denominator itself is a doubly robust and locally efficient estimator of $S_a(t \mid \tildelta = 1)$, but it was found to show worse performance than the $\widehat{S}^{\text{ACW}}_a(t)$ in a finite sample \citep{zhang2012contrasting,lee2022generalizable}.
A nonparametric bootstrap method can be used for a straightforward estimation of the variance of the ACW estimator.

\subsection{ACW with a Hazard Regression (HARE) Model}\label{sec:ACWhare}

The ACW estimator in \eqref{eq:ACW trans} is a consistent estimator under the PH assumption, which, in many case studies may not be correct.
In particular, as shown in Figure \ref{fig:surv ph viol} and the formal PH assumption tests, the PH assumption may be violated in the ACTG 175 data.% that we will analyze in Section \ref{sec:trans data}, 
The ACW estimator can be extended to account for such non-PH data using the HARE for the outcome model. 
The HARE model is based on the linear splines and their tensor products and does not depend on the PH assumption \citep{kooperberg1995hazard}. 

Let the vector of $p$-dimensional covariates $X$ lies in $\mathcal{X} \subseteq \mathbb{R}^p$ and let $\mathcal{G}$ be a $p$-dimensional linear space of functions on $[0,\infty) \times \mathcal{X}$ such that $g(\cdot \mid X)$ is bounded on $[0, \infty)$ for $g \in \mathcal{G}$ and $x \in \mathcal{X}$.  
The treatment-specific conditional log-hazard function for the HARE model is
\begin{align}
    \log \lambda_a^{H}(t \mid X) = \sum_{k=1}^p \beta^{H}_{ak} B_k(t\mid X) \label{eq:hare},
\end{align}
where $B_1(\cdot),\hdots, B_p(\cdot)$ are the basis of $\mathcal{G}$. The coefficients $\beta^{H}_{a} = (\beta^{H}_{a1},...,\beta^{H}_{ap})^T$ are estimated using the maximum likelihood estimation.
We use a superscript $H$ to denote the HARE model. 

The HARE model is implemented in the R function \textit{hare} in the R package \texttt{polspline}. It considers knots in the covariates and time and the pairwise interaction between covariates and time, and performs the variable selection simultaneously. For example, an $l$th knot of the $j$th covariate $x_j$ is represented by the basis function $(x_j - x_{jl})_+$, and an $l$th knot of time $t$ is represented by the basis function $(t_l - t)_+$, where $x_+ = \max(x, 0)$. Then the variable selection for the pairwise covariate interactions as well as covariates-time interactions are conducted using stepwise addition, stepwise deletion, and the Akaike Information Criterion.
If the interaction between the covariate and time is selected in the model, then the HARE model becomes a non-PH model.
We can easily combine the HARE model with the ACW estimator, by substituting $\widehat{\Lambda}_a^{H}(t \mid X_i) = \int_0^t \widehat{\lambda}_a^{H}(u \mid X_i) \mathrm{d}u$ in \eqref{eq:ACW trans} for $\widehat{\Lambda}_{ai}(u)$. With the HARE model, the ACW estimator gains robustness to the violation of the PH assumption.

\section{Results of Transporting the ACTG 175 Trial}\label{sec:trans data}

\subsection{Transporting the treatment effect to US early-stage HIV patients}\label{sec:US early}

The AIEDRP cohort study is an observational cohort study that enrolled patients within 1 year of having HIV antibody seroconversion \citep{hecht2006multicenter}. An important question could be an estimation of the effect of the combination therapy ZDV + ddI over ZDV monotherapy in the US early-stage HIV patients population represented by the AIEDRP study that consists of patients with higher CD4 count and less history of intravenous drug use, transporting from the ACTG 175 trial with intermediate-stage HIV patients. 

\begin{figure}[h]
\centerline{
\includegraphics[width=7in]{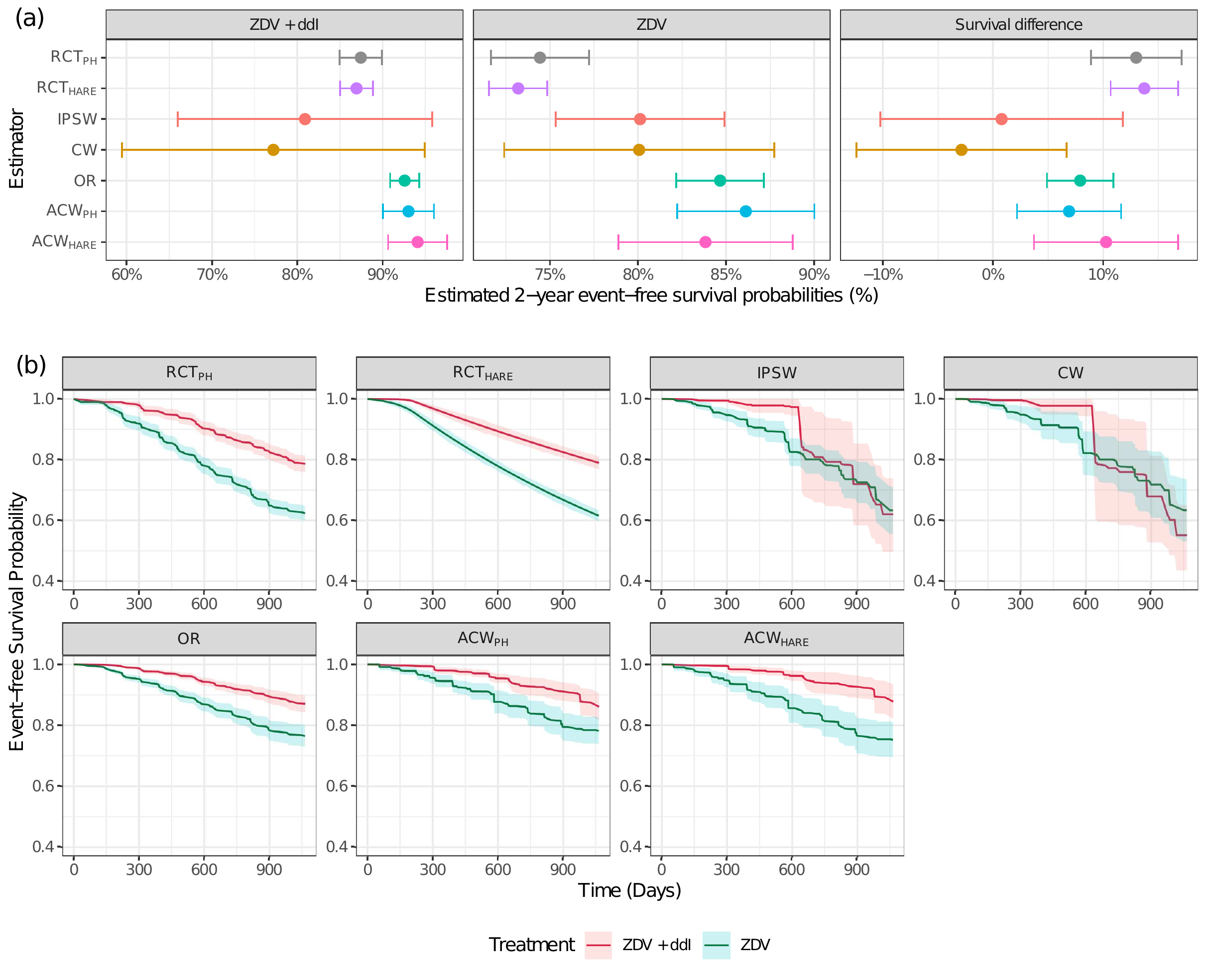}}
\caption{Estimated treatment-specific event-free survival probabilities in the US early-stage HIV patients; (a) 2-year event-free survival probabilities; (b) treatment-specific survival curves \label{fig:us early}}
\end{figure}

Figure \ref{fig:us early} plots estimated treatment-specific event-free survival probabilities in the US early-stage HIV patients. Figure \ref{fig:us early}(a)
%Table \ref{tab:us early} 
shows the estimated 2-year event-free survival probabilities by treatment and their differences in early-stage HIV patients in the US. 
The models based only on the ACTG 175 trial, i.e., $\rm{RCT_{PH}}$ and $\rm{RCT_{HARE}}$, give the estimated survival probabilities of about 84\% for the ZDV + ddI and 74\% for the ZDV. 
Thus, there is a 13\% higher chance of survival for ZDV + ddI over ZDV monotherapy, and no notable differences between $\rm{RCT_{PH}}$ and $\rm{RCT_{HARE}}$. 
On the other hand, after transporting the ACTG 175 trial, the estimated 2-year survival probabilities in the US early-stage HIV patients population are higher for both treatment groups in general. 
Specifically, the $\rm{OR_{PH}}$, $\rm{ACW}_{PH}$, and $\rm{ACW_{HARE}}$ estimators give an estimate of about 93\% and 84\% 2-year survival probabilities for ZDV + ddI and ZDV, respectively. 
The estimated 2-year survival differences in the early-stage HIV patients are smaller than that in ACTG 175 patients. 
The $\rm{OR_{PH}}$ and  $\rm{ACW}_{PH}$ estimators that are based on the Cox PH model show about a 7\% increase in 2-year survival probabilities.
On the other hand, the $\rm{ACW_{HARE}}$ estimator show about a 10\% increase in 2-year survival probability for ZDV + ddI over ZDV monotherapy. 
Given the possible violation of the PH assumption in the ACTG 175 trial, the estimated treatment effect under the PH model could be underestimated.
All these differences were significant at the 0.05 level.
Figure \ref{fig:us early}(b) plots the estimated treatment-specific survival curves.
After transporting to the US early-stage HIV patients, survival curves for both treatment groups are higher than the survival curves in the ACTG 175 trial patients across time. 
The combined therapy is more effective in treating the intermediate-stage HIV patients in the trial than early-stage patients. This may be because the combined therapy is more toxic and lower the compliance in the early-stage patients, which decreases the treatment effect of the combined therapy over the ZDV  monotherapy. %Thus, for early-stage patients, monotherapy may be considered.
%It can be seen that the estimated TATE with the HARE model shows a larger survival difference than those with the Cox PH model in early-stage patients.

The IPSW and CW estimators estimate treatment-specific survival curves that are in different shapes from the estimators with outcome models. Moreover, the estimated 2-year survival differences are not significant with larger variability. As IPSW and CW estimators are singly robust, these could be due to the misspecified sampling score model, resulting in the biased estimation of the treatment effect. In contrast, the ACW estimators are robust to such misspecification.

\clearpage

\subsection{Transporting the treatment effect to HIV patients in Thailand and southern Ethiopia}\label{sec:thai_and_etio}

Next, we transport the ACTG 175 trial to the HIV patients in two countries outside of the US: Thailand and southern Ethiopia. We first estimate the treatment effect in Thailand HIV patients. With about 500,000 HIV patients and 12,000 patients died of HIV-related diseases in 2020, Thailand is considered to be the country with a high HIV burden in Asia/Pacific regions \citep{manosuthi2021retrospective}. 
A retrospective study was conducted among the HIV patients registered in the national AIDS program database in eight provinces in Thailand. We are interested in evaluating the treatment effect in HIV patients in Thailand represented by the retrospective study. This observational study includes more than 10,000 patients with lower CD4 counts than the ACTG 175 patients. 

As mentioned in Section \ref{sec:motive}, the individual-level data for HIV patients in Thailand was not available. However, it is known that even only using summary statistics, the inferential population is well defined \citep{chu2022targeted}. 
We emulated the external observational data based on the baseline characteristics in Table \ref{tab:baseline}. Specifically, categorical covariates such as gender and CD4 category were generated from a categorical distribution with the given proportion. The numerical covariate age was generated from the shifted beta distribution, with mean and variance given from the summary data. The range of the beta distribution was determined following the ACTG 175 trial data unless otherwise specified in the original paper in Manosuthi et al.\citep{manosuthi2021retrospective}
To make the generated data more realistic, we incorporate the correlation structure of the ACTG 175 trial to the emulated Thailand data using the Gaussian copula, assuming that the correlation structure of the trial and the external data are similar. The sensitivity analyses show that the results are robust to the correlation structure as well as the randomness in data generation (see Appendix \ref{app:robustness}).

\begin{figure}[ht]
\centerline{
\includegraphics[width=7in]{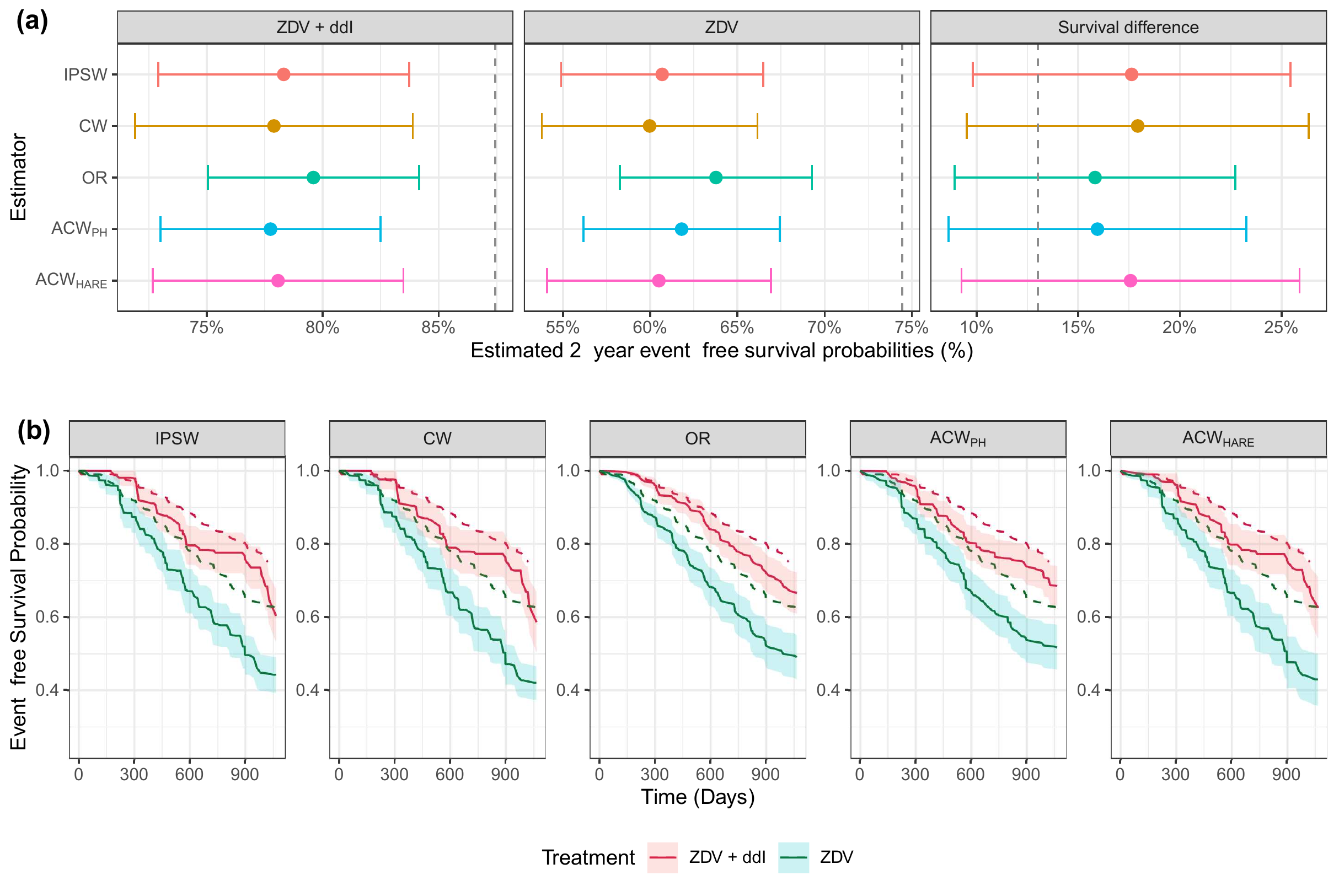}}
\caption{Estimated treatment-specific event-free survival probabilities in Thailand HIV patients; (a) 2-year event-free survival probabilities; (b) treatment-specific survival curves. Dashed lines indicate the estimates from the $\rm{RCT_{PH}}$ model. \label{fig:surv_thld}}
\end{figure}

Figure \ref{fig:surv_thld} depicts the estimated treatment-specific event-free survival probabilities in Thailand HIV patients.
Figure \ref{fig:surv_thld}(a) shows that 
all five estimators that transport the ACTG 175 trial results to the Thailand patients show about 16-17\% 2-year survival increase of ZDV + ddI over ZDV monotherapy, which is larger than the 13\% survival increase in the ACTG 175 trial patients shown in grey dashed vertical line. 
According to Figure~\ref{fig:surv_thld}(b), for both treatment groups, the estimated survival probabilities for Thailand HIV patients were found to be lower than that of the trial patients shown in dashed lines. 
These results make sense as the Thailand study consists of more than half of patients with low CD4 counts ($\le 200$), considered less healthy patients, than the ACTG 175 trial with only about 9\% patients with low CD4 counts. 
With the PH assumption, the estimated 2-year survival difference is around 15\%, whereas $\rm{RCT_{HARE}}$  shows a larger treatment effect, with about a 17.6\% survival probability increase. 
The latter is similar to the estimates from the IPSW and CW estimators that do not involve the PH assumption. In Figure \ref{fig:surv_thld}(b), estimated survival curves from the $\rm{RCT_{HARE}}$ estimator show different trends from the curves with the PH assumption and are shown to be more efficient than the estimated curves from the IPW-based estimators.

We are also interested in assessing the treatment effect in HIV patients in southern Ethiopia. A retrospective study was conducted including the HIV-infected patients enrolled in Dilla University Hospital located in southern Ethiopia \citep{hailemariam2016determinants}. The study consists of more females and patients having lower CD4 count compared to the ACTG 175 patients. As the individual-level data for southern Ethiopia patients data is not available, we emulate the external observational data based on the summary statistics in Table \ref{tab:baseline} and the correlation structure of the ACTG 175 data, similar to the Thailand data.

\begin{figure}[ht]
\centerline{
\includegraphics[width=7in]{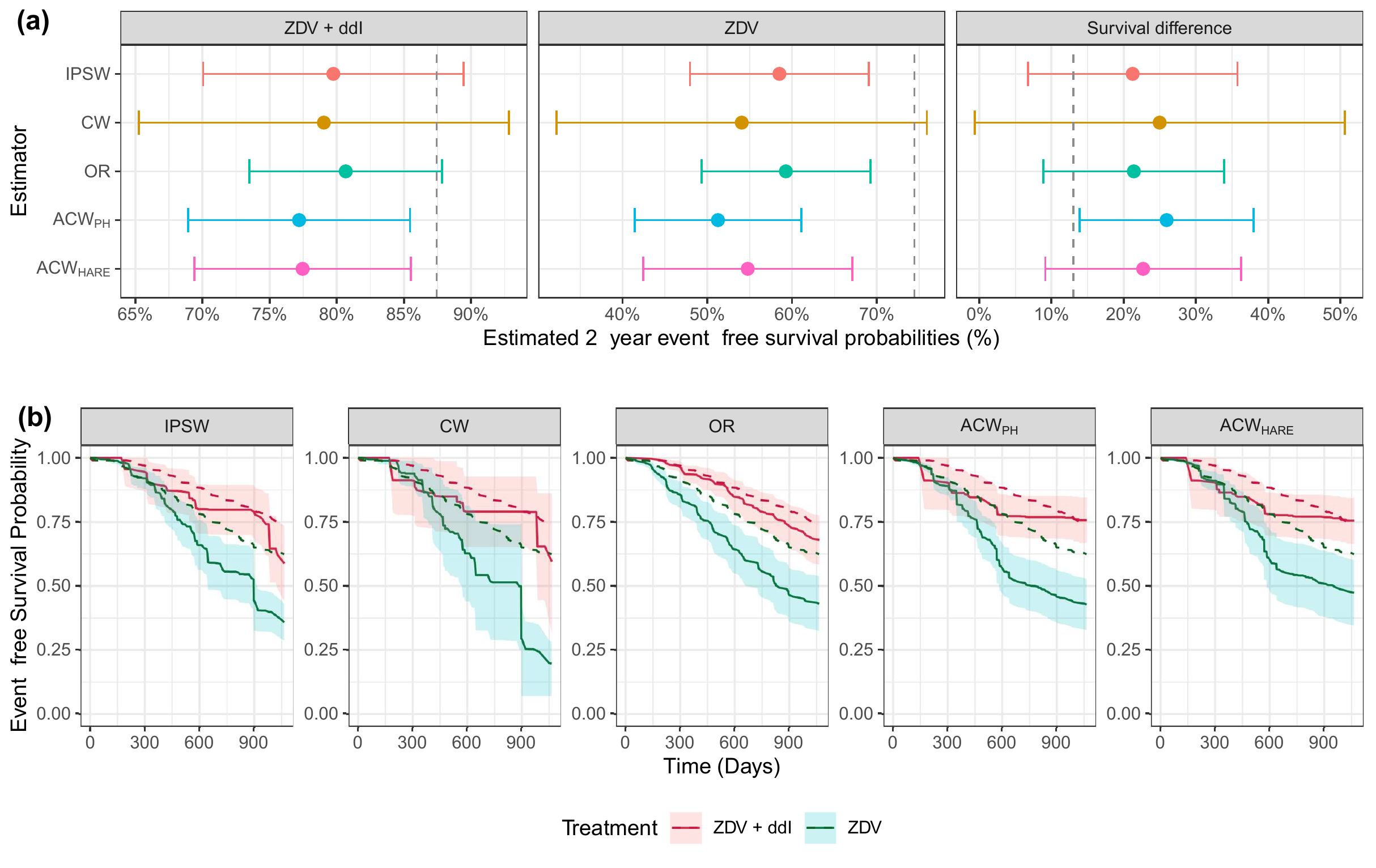}}
\caption{Estimated treatment-specific event-free survival probabilities in southern Ethiopia HIV patients; (a) 2-year event-free survival probabilities; (b) treatment-specific survival curves. Dashed lines indicate the estimates from the $\rm{RCT_{PH}}$ model. \label{fig:surv_etio}}
\end{figure}

Figure \ref{fig:surv_etio} plots the estimated treatment-specific event-free survival probabilities in southern Ethiopia HIV patients. 
Figure \ref{fig:surv_etio}(a) shows the estimated 2-year survival difference after transporting the ACTG 175 trial to the southern Ethiopia patients. The estimated ATE in southern Ethiopia patients is larger than that in the ACTG 175 trial in a grey dashed vertical line. 
Figure \ref{fig:surv_etio}(b) shows estimated treatment-specific survival curves. All five curves show a trend of no significant difference at the initiation of the treatment and then significance after a year or more on treatment. The $\rm{RCT_{PH}}$ and $\rm{RCT_{HARE}}$ show similar estimated survival curves. The IPSW and CW estimator show relatively large variability, which could be due to the misspecification of the sampling score model.

\section{Conclusions}\label{sec:conclude}

In this paper, we have explored various models for transporting the treatment effect from the ACTG 175 trial to the three target populations including both the PH and non-PH models. 
Due to the heterogeneity in the patient characteristics between the ACTG 175 trial and the external data that represent the target populations, the TATE is different from the treatment effect estimated in the trial. 
Specifically, as patients in the US early-stage HIV patients are relatively healthier than the patients in ACTG 175 trial who are in intermediate-stage disease, the estimated treatment-specific survival probabilities are higher and their difference is smaller in the target population compared to those in the trial. 
On the other hand, as HIV patients in Thailand and southern Ethiopia are relatively sicker with lower baseline CD4 counts than the ACTG 175 trial patients, the estimated treatment-specific survival probabilities were found to be lower and differences were larger in general. 
These results correspond to that of Kennedy et al. \citep{kennedy2021semiparametric} in that combination therapy could benefit more  the high-risk or sicker patients.
Moreover, the estimated TATEs from the ACW with the HARE model and the ACW with the Cox PH model
are different. 
As the PH assumption may be violated in the ACTG 175 trial, the latter might be biased whereas the former corrects the bias by accounting for the non-PH model. 
The ACW with the HARE model also shows more robustness and efficiency in general than the IPSW and CW estimators which do not depend on the PH assumption but are singly robust to the sampling score model.

The presented analysis is based on the assumption that the trial participation is ignorable, i.e., all covariates related to the trial participation and survival time are captured. 
Even though some covariates that are known to be highly predictive of the disease progression were captured in the external observational data, some important covariates may not be available as the observational samples were not originally collected for the research purpose. 
Sensitivity analyses can be conducted to assess the robustness of the presented results in the presence of unmeasured covariates in external data, e.g., VanderWeele and Ding \citep{vanderweele2017sensitivity} and Yang and Lok \citep{yang2017sensitivity}. 
Moreover, some external data, such as Thailand or southern Ethiopia studies, do not provide individual-level data but only the summary statistics of the covariate distribution from the target populations. 
Several recent works suggest that the analysis based on reliable summary statistics for covariates could lead to the valid inference of the target population, e.g., Chu et al \citep{chu2022targeted}.
Also, the proposed external data generating process
was found to be robust to the correlation structure and the randomness in data generation.
Nonetheless, the results from the emulated data could be different from the results analyzing the actual individual-level data.

 \bibliographystyle{Chicago}
\bibliography{ref}

%%%%%%%%%%%%%%%%%%%%%%%% Appendix %%%%%%%%%%%%%%%%%%%%%%%

\newpage{} 
\begin{center}
\textbf{\Large{}Appendix}{\Large{} }{\Large\par}
\par\end{center}

%\pagenumbering{arabic} %reset page counter to 1
%\renewcommand*{\thepage}{A\arabic{page}}

\setcounter{section}{0} 
\global\long\def\thesection{\Alph{section}}%
\setcounter{equation}{0}
\global\long\def\thesubsection{\Alph{section}.\arabic{subsection}}%
\setcounter{equation}{0} 
\global\long\def\theequation{\Alph{section}\arabic{equation}}%
\setcounter{figure}{0} 
\global\long\def\thefigure{\Alph{section}\arabic{figure}}%
\setcounter{table}{0} 
\global\long\def\thetable{\Alph{section}\arabic{table}}%

\section{Robustness of the individual-level data generation process in Thailand and southern Ethiopia}
\label{app:robustness}
In this section, we explore the robustness of the individual-level data generation in Thailand and southern Ethiopia based on the summary statistics in Table \ref{tab:baseline}.

\subsection{Under the increased correlation between age and CD4 count\label{app1.1a}}
For the ACTG 175 trial, the Pearson's correlation between age and CD4 count is -0.043 and between the age and the CD4 category is -0.04. 
Assuming that the trial data and the external data have the same correlation structure, we use these values to emulate the external data in Section \ref{sec:thai_and_etio}. 
However, the correlation in the trial and the external data could be different; 
it is reasonable to assume that age and CD4 count are highly correlated as both the elderly and lower CD4 counts imply sicker patients in general.
Therefore, we investigate the effect of increasing the correlation between age and CD4 count.
Specifically, we emulated the external observational data with an increased correlation of -0.8 instead of - 0.04 between the age and CD4 category for Thailand HIV patients.
Similarly, for southern Ethiopia HIV patients, we emulated the individual-level data with a correlation of -0.8 between the age and CD4 count.

\begin{figure}[ht]
\centerline{
\includegraphics[width=7in]{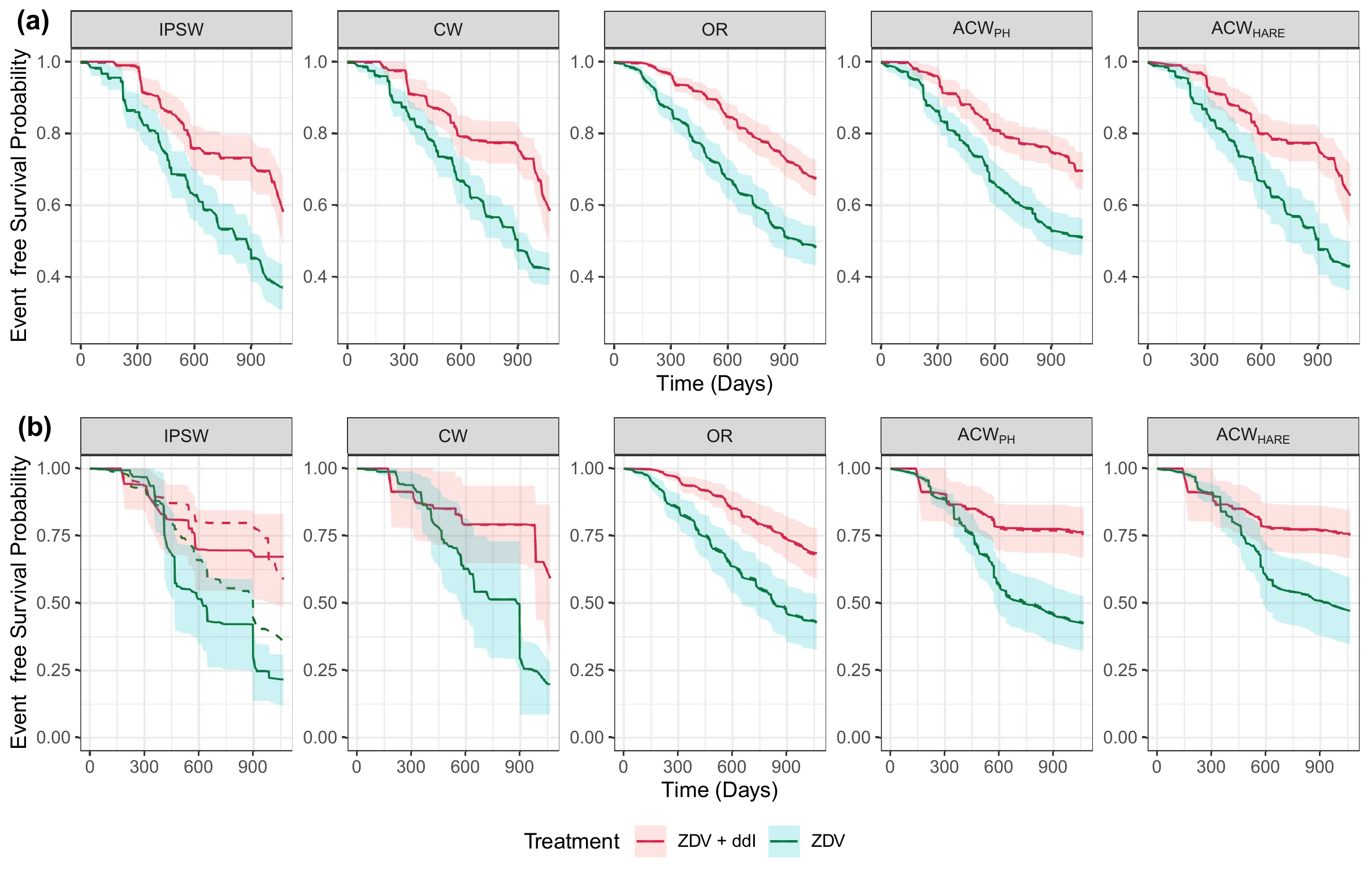}}
\caption{Estimated treatment-specific event-free survival curves with increased correlation between age and CD4 count/category; (a) estimated survival curves in Thailand HIV patients; (b) estimated survival curves in southern Ethiopia HIV patients. \label{fig:surv corr}}
\end{figure}

Figure \ref{fig:surv corr} depicts the estimated treatment-specific survival curves with the increased correlation of -0.8 between age and CD4 count/category. 
Dashed lines indicate the estimated treatment-specific survival curves based on the correlation structure in the ACTG 175 trial. 
Figure \ref{fig:surv corr}(a) shows that all five transport methods give an almost identical estimation of survival curves under the trial correlation of -0.05 and increased correlation of -0.8 for Thailand HIV patients.
Similarly, Figure \ref{fig:surv corr}(b) shows that, except the IPSW estimator, the estimated survival curves for southern Ethiopia HIV patients under the highly negative age-CD4 count correlation are almost the same as that under the correlation in the trial data. For the IPSW estimator, the discrepancy between the different correlation structures may be due to its intrinsic weakness that it is unstable under the extreme sampling score.

\subsection{Robust to randomness in data generation \label{app1.1.1a}}

\begin{figure}[h]
\centerline{
\includegraphics[scale = 0.6]{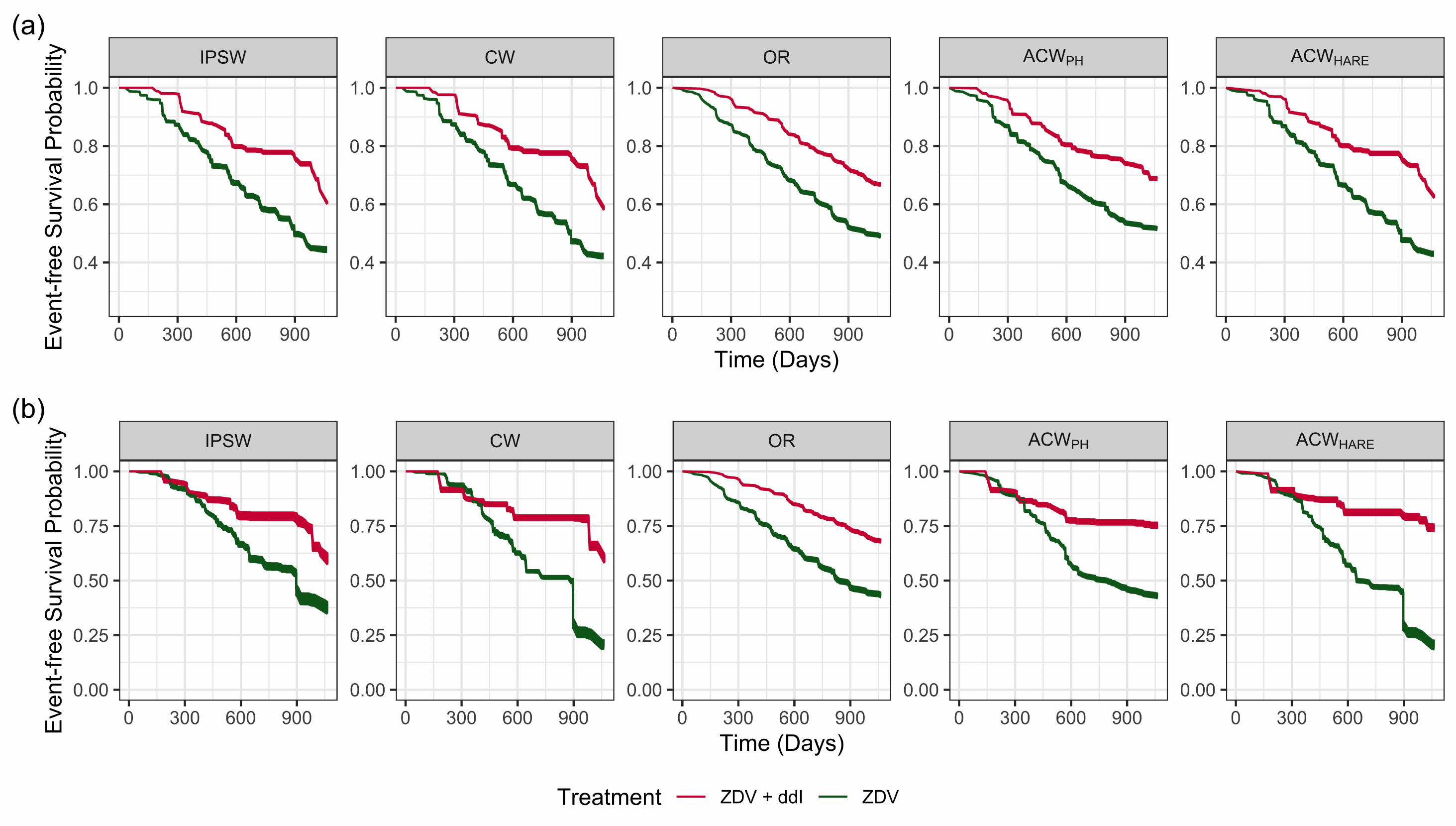}}
\caption{Estimated treatment-specific event-free survival curves from 1,000 sets of emulated data; (a) Thailand; (b) southern Ethiopia. \label{fig:repeat robust}}
\end{figure}

We now show that the external data generation process we used is robust to the randomness in data generation. Figure \ref{fig:repeat robust} plots the estimated treatment-specific event-free survival curves from 1,000 sets of emulated data based on the process described in Section \ref{sec:thai_and_etio} in the main text. Specifically, categorical covariates were
generated from a categorical distribution based on the summary proportion, and the numerical covariates were generated from the shifted beta distribution with mean and standard deviation from the summary statistics. Even though the 1,000 emulated datasets were different from each other due to the randomness in the probability distribution, it can be seen that for all five transport methods, the estimated curves were almost identical for both Thailand and southern Ethiopia.

% \begin{figure}[h]
% \centerline{
% \includegraphics[width=7in]{pics/etio_repeat.jpeg}}
% \caption{Robust Ethiopia \label{fig:repeat etio}}
% \end{figure}

\section{Effect of two other treatments - ZDV + ZAL and ddI - over ZDV}
\label{app:other trts}

In the ACTG 175 trial, enrolled patients were randomly assigned to the four treatments, ZDV + ddI, ZDV + ZAL, ddI, and ZDV.
In the main text, we focus on the effect of ZDV + ddI combination therapy over ZDV for illustrative purpose. This section provides the results of transporting the effect of two other treatments, i.e., ZDV + ZAL and ddI, over ZDV to the external target populations.

\subsection{Transporting the effect of ZDV + ZAL over ZDV}
\label{app:plus zal}

In this section, we consider ZDV + ZAL and ZDV as binary treatment, which consists of 524 ZDV + ZAL patients and 532 ZDV monotherapy patients. Similar to the analysis in the main text, the primary endpoint is the progression of HIV disease, defined as a more than 50 percent decline in the CD4 cell count or development of the acquired immunodeficiency syndrome, or death. The causal estimand of interest is a 2-year event-free survival difference between ZDV + ZAL and ZDV monotherapy, and about 73\% of the survival times were right-censored.

\begin{figure}[ht]
\centerline{
\includegraphics[width=7in]{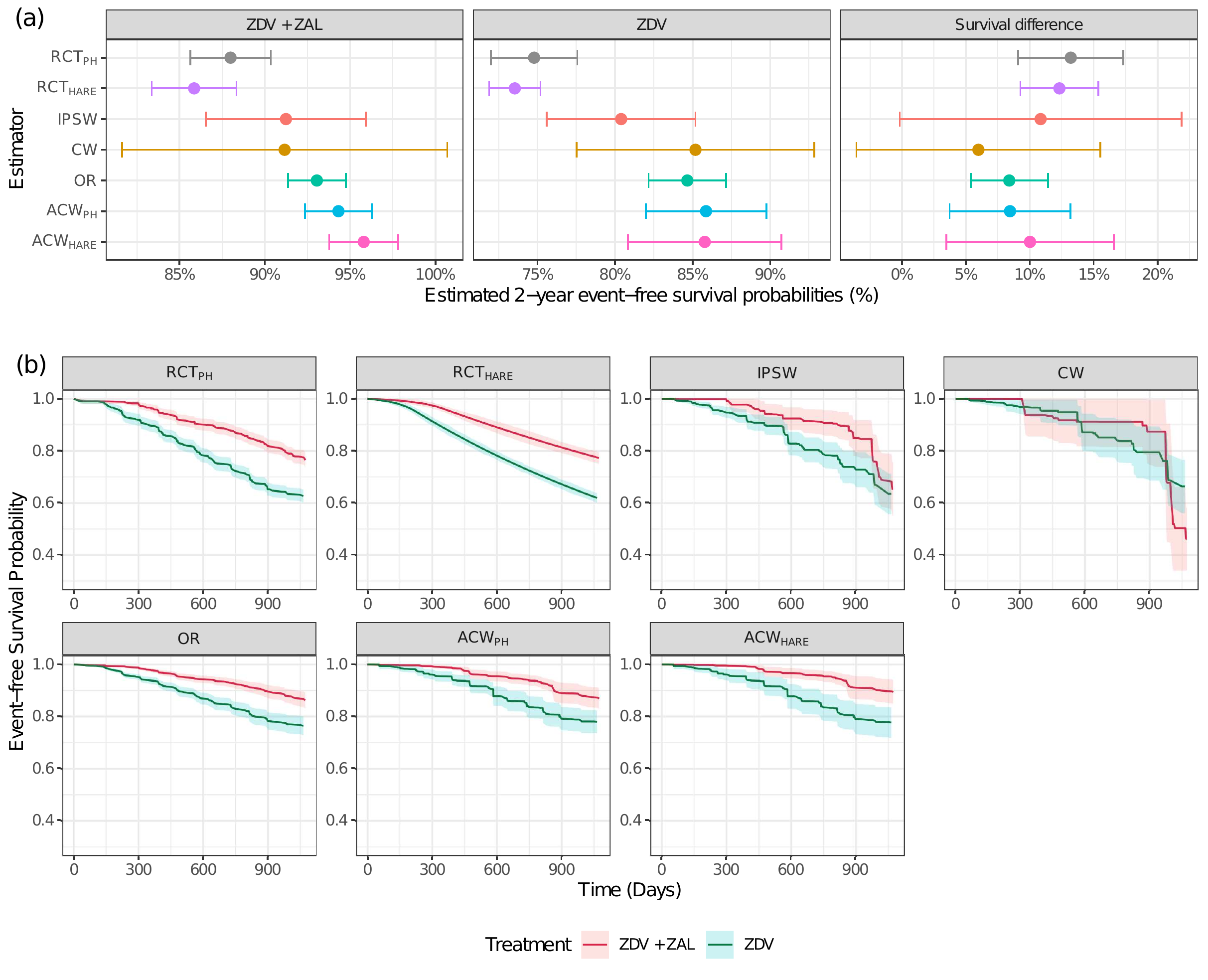}}
\caption{Estimated treatment-specific event-free survival probabilities in the US early-stage HIV patients for ZDV + ZAL vs. ZDV; (a) 2-year event-free survival probabilities; (b) treatment-specific survival curves \label{fig:us early plus zal}}
\end{figure}

Figure \ref{fig:us early plus zal} plots estimated treatment-specific event-free survival probabilities in the US early-stage HIV patients. 
Figure \ref{fig:us early plus zal}(a) shows that the transported 2-year survival probabilities in the US early-stage HIV patients population are higher for both treatment groups, and their differences are smaller in general than those estimated from the ACTG 175 trial.
For instance, the estimators based only on the trial data, i.e., $\rm{RCT_{PH}}$ and $\rm{RCT_{HARE}}$, give an estimate of a 13\% higher survival probability for ZDV + ZAL over ZDV monotherapy.
On the other hand, the transport methods estimate that the 2-year survival differences in the US early-stage HIV patients are 8\%-10\%.
Figure \ref{fig:us early plus zal}(b) illustrates that after transporting to the US early-stage HIV patients, survival curves for both treatment groups are higher than those in the ACTG 175 trial patients across time. 
These results suggest that as the US early-stage patients are healthier than the ACTG-175 trial participants, the effect of the  combined therapy over the ZDV monotherapy is less significant in the target population, possibly due to the toxicity and the low compliance.
%comply to that of ZDV + ddI vs. ZDV in the main text.
There is no notable difference between the estimated survival probabilities from the $\rm{ACW_{HARE}}$ estimator and the estimators with the PH assumption, i.e., $\rm{OR_{PH}}$, $\rm{ACW}_{PH}$.%, implying that the PH assumption may not be violated.

\begin{figure}[ht]
\centerline{
\includegraphics[width=7in]{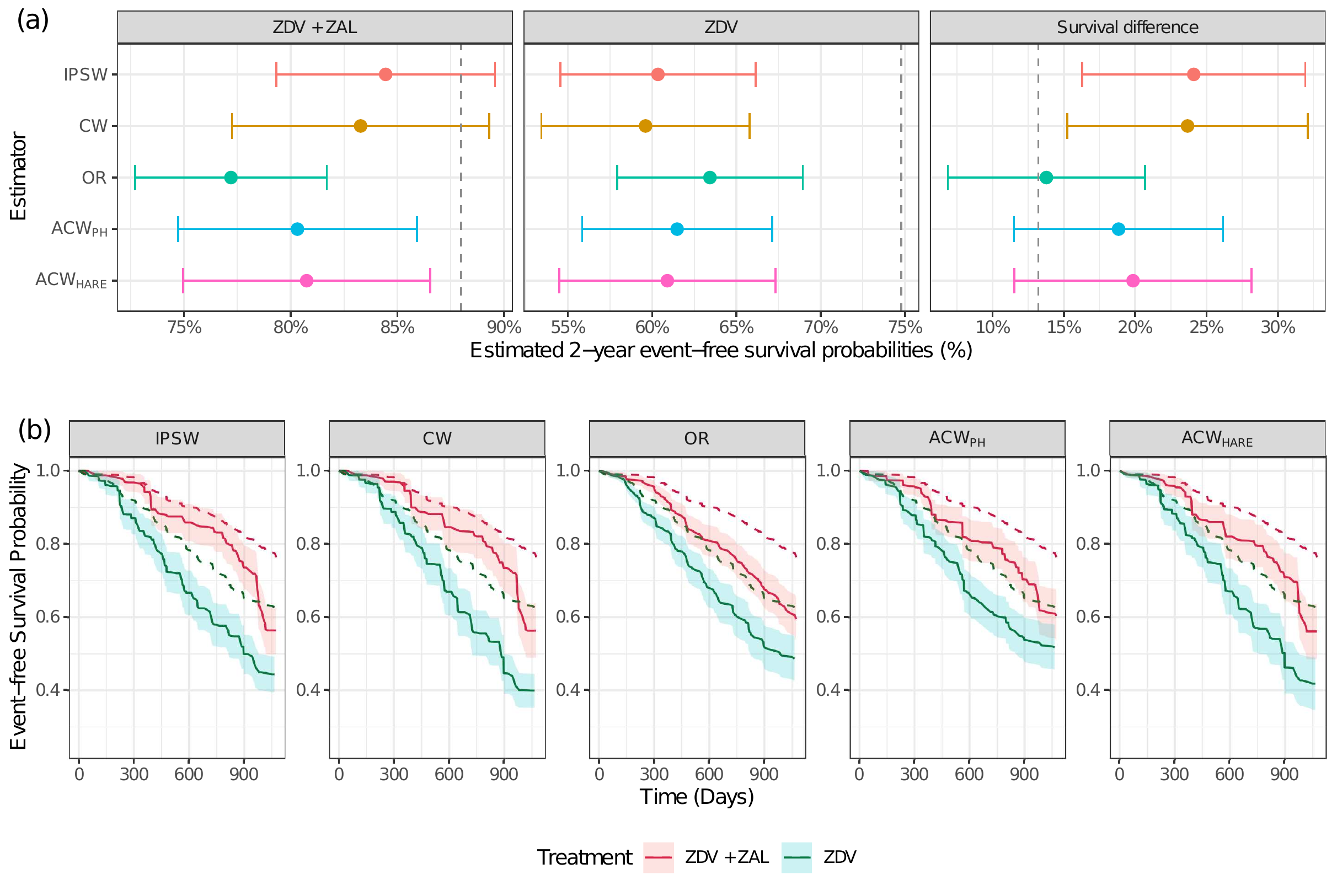}}
\caption{Estimated treatment-specific event-free survival probabilities in Thailand HIV patients for ZDV + ZAL vs. ZDV; (a) 2-year event-free survival probabilities; (b) treatment-specific survival curves. Dashed lines indicate the estimates from the $\rm{RCT_{PH}}$ model. \label{fig:surv_thld_plus_zal}}
\end{figure}

\begin{figure}[ht]
\centerline{
\includegraphics[width=7in]{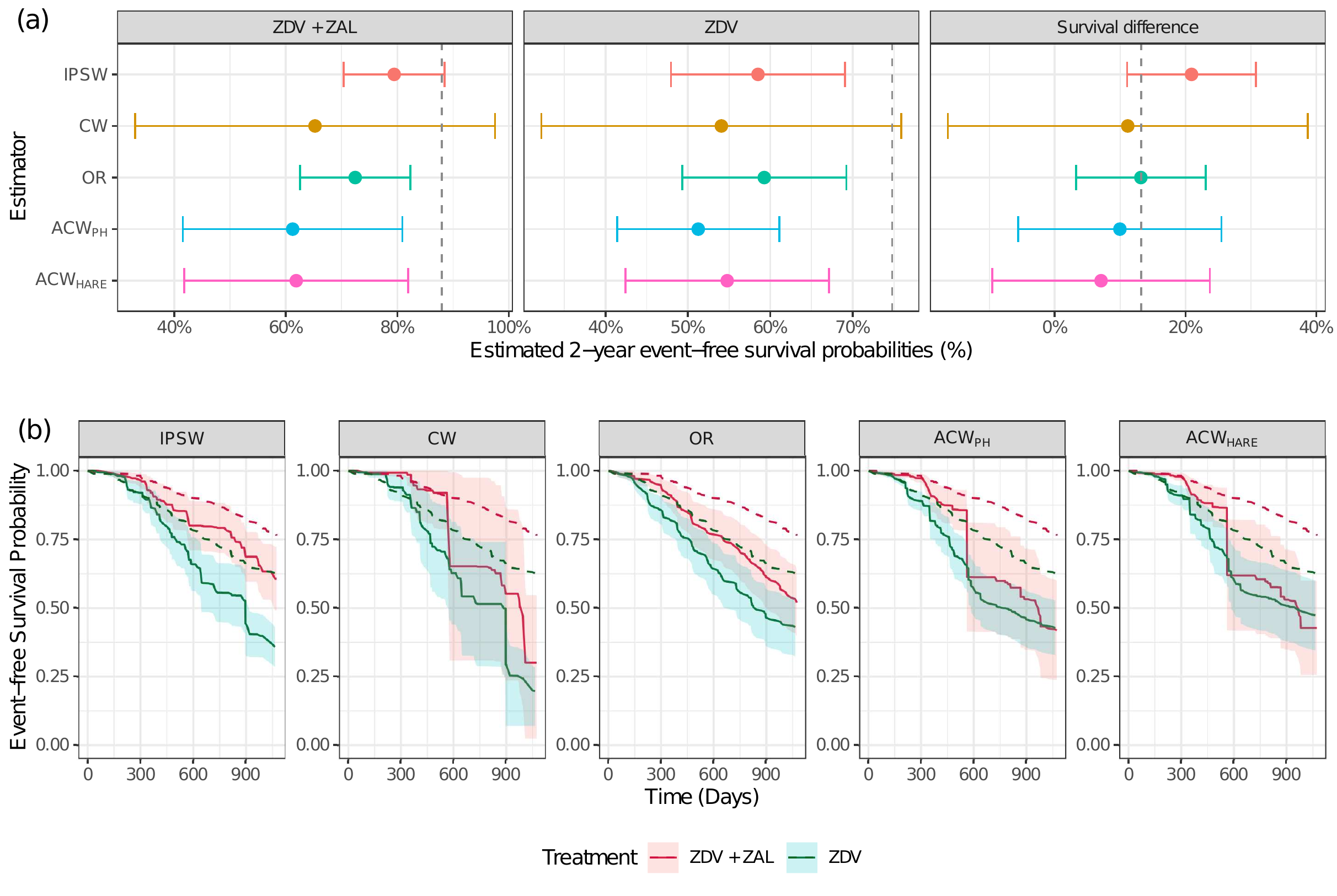}}
\caption{Estimated treatment-specific event-free survival probabilities in southern Ethiopia HIV patients for ZDV + ZAL vs. ZDV; (a) 2-year event-free survival probabilities; (b) treatment-specific survival curves. Dashed lines indicate the estimates from the $\rm{RCT_{PH}}$ model. \label{fig:surv_etio_plus_zal}}
\end{figure}

For Thailand and southern Ethiopia HIV patients, we emulated the external data using the same data generation process described in the main text. Figure \ref{fig:surv_thld_plus_zal} and Figure \ref{fig:surv_etio_plus_zal} depict the estimated survival probabilities for the ZDV + ZAL and ZDV treatment groups in Thailand and southern Ethiopia HIV patients, respectively. As patients in both target populations are considered to be sicker than the patients in the ACTG 175 trial, the transported treatment-specific survival curves are lower than the survival curves estimated from the trial data, depicted in dashed lines. According to Figure  \ref{fig:surv_thld_plus_zal}(b), it can be seen that the estimated survival curves by $\rm{ACW_{HARE}}$ are steeper than those by the $\rm{OR_{PH}}$ and $\rm{ACW}_{PH}$ estimators which depend on the PH assumption. The former estimator is robust to the violation of the PH assumption whereas the latter estimators are not, which may result in the overestimation of the survival probabilities.
Figure \ref{fig:surv_etio_plus_zal}(b) shows that there is no significant effect of the ZDV + ZAL combined therapy over ZDV in  southern Ethiopia HIV patients.
However, this result may not be meaningful due to the large variability of the transported probabilities.

\newpage
\subsection{Transporting the effect of ddI over ZDV}
This section considers ddI and ZDV as binary treatment, consisting of 561 ddI patients and 532 ZDV monotherapy patients. The analyses are similar to those of ZDV + ddI vs. ZDV in the main text and ZDV + ZAL vs. ZDV in Appendix \ref{app:plus zal}. About 72\% of the survival times were right-censored.

\begin{figure}[h]
\centerline{
\includegraphics[width=7in]{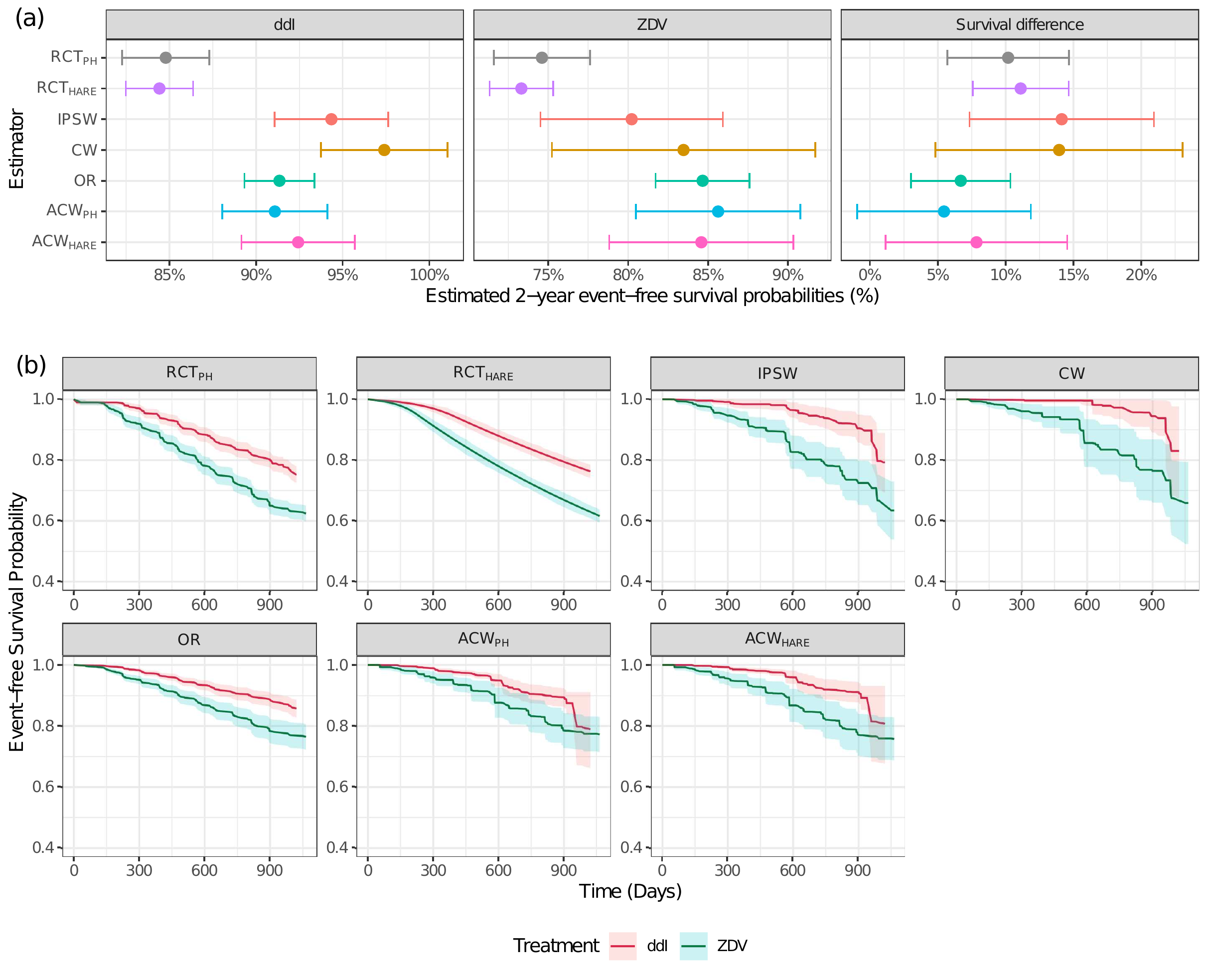}}
\caption{Estimated treatment-specific event-free survival probabilities in the US early-stage HIV patients for ddI vs. ZDV; (a) 2-year event-free survival probabilities; (b) treatment-specific survival curves \label{fig:us early vs ddi}}
\end{figure}

Figure \ref{fig:us early vs ddi} shows that the estimated survival probabilities for both the ddI and ZDV treatment groups in the US early-stage HIV patients population are higher than those in the ACTG 175 trial as the former is healthier. The OR,  $\rm{RCT_{PH}}$, and $\rm{RCT_{HARE}}$ estimators suggest that the effect of ddI over ZDV is smaller in the target population. 
This result contradicts the estimated survival probabilities from the IPSW and CW estimators which is larger than the one in the trial data. The estimated survival probabilities from the $\rm{ACW_{HARE}}$ estimator and that from the estimators with the PH assumption, i.e., $\rm{OR_{PH}}$, $\rm{ACW}_{PH}$, were found to be similar.

\begin{figure}[ht]
\centerline{
\includegraphics[width=7in]{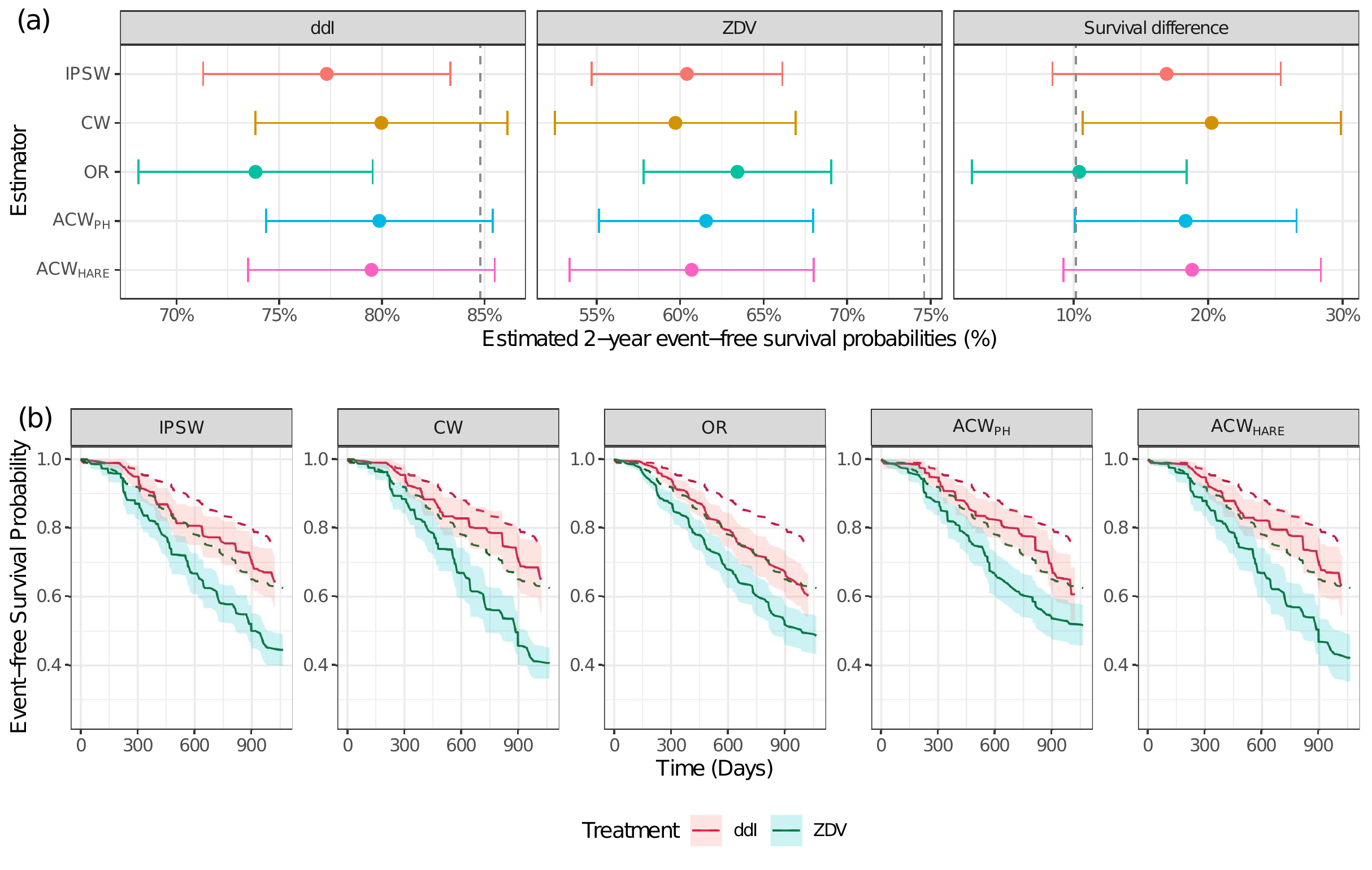}}
\caption{Estimated treatment-specific event-free survival probabilities in Thailand HIV patients for ddI vs. ZDV; (a) 2-year event-free survival probabilities; (b) treatment-specific survival curves. Dashed lines indicate the estimates from the $\rm{RCT_{PH}}$ model. \label{fig:surv_thld_vs_ddi}}
\end{figure}

\begin{figure}[ht]
\centerline{
\includegraphics[width=7in]{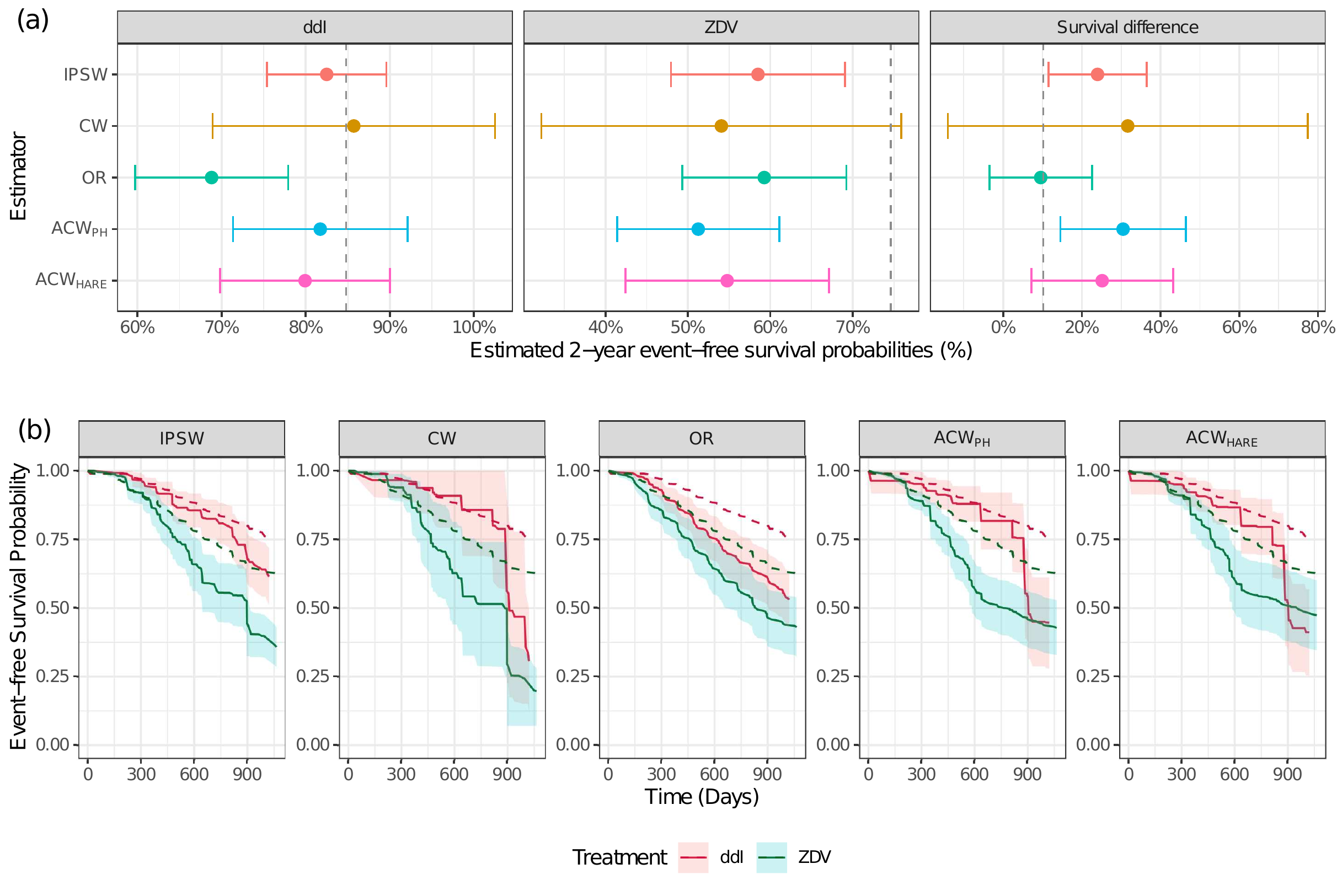}}
\caption{Estimated treatment-specific event-free survival probabilities in southern Ethiopia HIV patients for ddI vs. ZDV; (a) 2-year event-free survival probabilities; (b) treatment-specific survival curves. Dashed lines indicate the estimates from the $\rm{RCT_{PH}}$ model. \label{fig:surv_etio_vs_ddi}}
\end{figure}

The transported results of ddI vs. ZDV for Thailand and southern Ethiopia HIV patients are parallel to those of other treatment sets; the estimated treatment-specific survival curves in the target populations are lower than the estimated survival curves in the trial data as the target populations include less healthy patients. Under the same external data generation process described in the main text, Figure \ref{fig:surv_thld_vs_ddi} and Figure \ref{fig:surv_etio_vs_ddi} show the estimated survival probabilities for the ddI and ZDV treatment groups in Thailand and southern Ethiopia HIV patients, respectively, which are lower than the survival curves estimated from the trial data, depicted in dashed lines. 
According to Figure \ref{fig:surv_thld_vs_ddi}(b), the estimated survival curves by $\rm{ACW_{HARE}}$ were found to be steeper than those by the $\rm{OR_{PH}}$ and $\rm{ACW}_{PH}$ estimators which depend on the PH assumption. As the $\rm{OR_{PH}}$ and $\rm{ACW}_{PH}$ estimators can not adjust for the violation of the PH assumption, the survival probabilities might be overestimated.
In Figure \ref{fig:surv_etio_vs_ddi}(a), all estimators except the OR estimator estimate the transported 2-year survival difference in the southern Ethiopia HIV patients to be higher than that in the trial data. According to Figure \ref{fig:surv_etio_vs_ddi}(b), all transport methods besides the OR estimator show a delayed treatment effect and crossing of the transported survival curves for ddI and ZDV.

\end{document}